\documentclass[aps,showpacs,preprintnumbers,amsmath,amssymb,eqsecnum]{revtex4}

\usepackage{graphicx}
\usepackage{dcolumn}
\usepackage{bm}

\begin{document}

\newcommand{\balpha}{\bm{\alpha}}
\newcommand{\bmu}{\bm{\mu}}
\newcommand{\bsig}{\bm{\sigma}}
\newcommand{\brho}{\bm{\rho}}
\newcommand{\blam}{\bm{\lambda}}
\newcommand{\bL}{\bm{L}}
\newcommand{\bA}{\bm{A}}
\newcommand{\bB}{\bm{B}}
\newcommand{\bC}{\bm{C}}
\newcommand{\bS}{\bm{S}}
\newcommand{\bx}{\bm{x}}
\newcommand{\by}{\bm{y}}
\newcommand{\bz}{\bm{z}}
\newcommand{\bk}{\bm{k}}
\newcommand{\bv}{\bm{v}}
\newcommand{\mb}{\textbf m}
\newcommand{\pb}{\textbf p}
\newcommand{\rb}{\textbf r}
\newcommand{\sbd}{\textbf s}

\title{Electromagnetic corrections to baryon masses}

\author{Loyal Durand}%
 \email{ldurand@hep.wisc.edu}
\affiliation{%
Department of Physics, University of Wisconsin-Madison \\
             Madison, Wisconsin 53706, USA
}%

\author{Phuoc Ha}%
\email{pdha@iusb.edu}
\affiliation{%
Department of Physics and Astronomy, Indiana University South Bend \\
             South Bend IN 46634, USA \\
             and \\
Institute of Physics and Electronics,
Vietnamese Academy of Science and Technology \\
10 Dao Tan, Ba Dinh, Hanoi, Vietnam
}%

\date{\today}

\begin{abstract}

We analyze the electromagnetic contributions to the octet and decuplet baryon masses using the heavy baryon approximation in chiral effective field theory and methods we developed in earlier analyses of the baryon masses and magnetic moments. Our methods connect simply to Morpurgo's general parametrization of the electromagnetic contributions and to semirelativistic quark models. Our calculations are carried out including the one-loop mesonic corrections to the basic electromagnetic interactions, so to two loops overall. We find that to this order in the chiral loop expansion there are no three-body contributions. The Coleman-Glashow relation and other sum rules derived in quark models with only two-body terms therefore continue to hold, and violations involve at least three-loop processes and can be expected to be quite small. We present the complete formal results and some estimates of the matrix elements here. Numerical calculations will be presented separately.

\end{abstract}


\maketitle


\section{\label{sec:introduction} Introduction}

Electromagnetic corrections to baryon masses have been of interest for many years. Early attempts to calculate the neutron-proton mass difference as arising from the electromagnetic self energies were largely unsuccessful.  This history can be traced starting, for example, from \cite{Zee} and the references therein. Progress on the modern theory began with the introduction of flavor SU(3) symmetry and the derivation by Coleman and Glashow \cite{Coleman-Glashow} of the electromagnetic sum rule
\begin{equation}
\label{CG}
( \Xi^--\Xi^0)= (\Sigma^--\Sigma^+)-(n-p),
 \end{equation}
where we denote the particle masses by the particle names. This sum rule is remarkably accurate with a difference between the two sides  that is zero to within about one standard deviation, $l-r=-0.31\pm 0.25$ MeV \cite{PDG}.  This is small on the scale of the SU(3) symmetry breaking in the mass spectrum and also compared to the individual mass differences. For example, the difference $(\Xi^--\Xi^0)$ is $6.74\pm 0.02$ MeV, some twenty times larger.
 
 It was later understood on the basis of the  nonrelativistic quark model \cite{Rubenstein1,Rubenstein2,Ishida,Gal-Scheck} that the Coleman-Glashow relation is actually independent of the breaking of the SU(3) symmetry and holds exactly along with several other sum rules in the absence of any three-body interactions among the quarks in the baryons. The results are consistent with the structure of the interactions among quarks expected in QCD-based quark models \cite{DeRujula,Rosner}.   
 
The most general expression for the purely electromagnetic contributions to the mass differences in the baryon octet and decuplet has since been determined by Morpurgo \cite{Morpurgo_EM1} using his general parametrization method for amplitudes is QCD. The results were again shown to satisfy the known sum rules for mass splittings within isospin multiplets independently of the symmetry breaking provided there are no terms with three flavor labels. The sum rules have also been analyzed in the $1/N_c$ expansion of QCD by Jenkins and Lebed \cite{Jenkins4}, who give definite predictions for the expected sizes of any deviations.  
 
The interest in the electromagnetic mass splittings and sum rules has been renewed recently \cite{Rosner,Morpurgo_EM2,Jenkins5} with improvements in the measurements of the $\Xi^0$ mass \cite{PDG} and in the accuracy with which the Coleman-Glashow relation, other sum rules, and the $1/N_c$ expansion can now be tested. It would also be of interest to actually calculate rather than just parametrize the  the electromagnetic contributions to the octet and decuplet masses.  

In the present paper, we analyze these contributions using the heavy baryon approximation in chiral effective field theory and methods developed in our earlier analyses of the baryon masses \cite{DHJ1,DHJ2,DH_masses} and magnetic moments \cite{DHJ2,DH_moments3}. Our methods connect simply to Morpurgo's parametrization and to semirelativistic quark models. The calculations are carried out including the one-loop mesonic corrections to the basic electromagnetic interactions, so to two loops overall. We find that to this order in the chiral loop expansion there are no three-body contributions. The original two-body sum rules therefore still hold, and violations involve at least three-loop processes and can be expected to be quite small. We present the complete formal results and some estimates of the matrix elements here. Numerical calculations will be presented separately.


\section{\label{sec:background}Background}


\subsection{\label{subsec:parametrization} Parametrization of electromagnetic corrections to baryon masses}

Morpurgo \cite{Morpurgo_EM1} has given a  parametrization for the electromagnetic contributions to the  masses of the ordinary octet and decuplet baryons at order $e^2$ based on his general parametrization method 
\cite{Morpurgo_param} for amplitudes in QCD. In this method, the exact states of the system are written in terms of the action of unitary operators on model states which have the structure of the baryons in the nonrelativistic quark model. The model states are completely labelled by their spin and flavor structure in terms of the constituent quarks. As a result, the parametrization of an arbitrary one-baryon matrix element can depend only on the those labels and can be related to matrix elements of a set of independent spin- and  flavor-dependent operators $\Gamma_i$ in the quark model states.  

In the case of $\textrm{O}(e^2)$ contributions to the baryon masses, the  $\Gamma$'s must be bilinear in the quark charge matrix $Q = \textrm{diag}\,(2/3,-1/3,-1/3)$ and can depend otherwise on the quark spin matrices $\bsig$ and and flavors. Ignoring isospin breaking through the small $u$, $d$ mass difference, Morpurgo groups the $\Gamma$'s in \cite{Morpurgo_EM1} according to the numbers of strange-quark projection operators that appear, representing the degree of symmetry breaking through the strange quark mass. We will denote that operator, called $P^\lambda$ in  \cite{Morpurgo_EM1}, by $M^s=\textrm{diag}\,(0,0,1)$.\footnote{$M^s$ was denoted by $M$ in \cite{DHJ1,DHJ2,DH_moments3} where the light-quark mass differences did not play a role.} Using the conventions that
\begin{equation}
\label{sums}
\sum[i]\equiv\sum_{i},\quad \sum[ij]\equiv\frac{1}{2}\sum_{i\not=j},\quad \sum[ijk]\equiv\frac{1}{6} \sum_{i\not=j\not=k}, 
\end{equation}
where $ i,j,k\in u,d,s$ label the three quarks in a baryon, we can regroup Morpurgo's results into sets of one-, two, and three-quark operators as follows:
\begin{eqnarray}
&& \textrm{One-body operators:} \nonumber \\
\label{1,7}
&& \Gamma_1=\sum[Q_i^2],\quad \Gamma_7=\sum[Q_i^2M^s_i]. 
\end{eqnarray}
\begin{eqnarray}
&& \textrm{Two-body operators:} \nonumber \\
\label{2,4,5}
&& \Gamma_2=\sum[Q_i^2(\bsig_i\cdot\bsig_j)], \quad \Gamma_4=\sum[Q_iQ_j],\quad \Gamma_5 = \sum[Q_iQ_j(\bsig_i\cdot\bsig_j)], \\
\label{8,10,11}
&& \Gamma_8=\sum[Q_i^2M^s_i(\bsig_i\cdot\bsig_j)] ,\quad \Gamma_{10}=\sum[Q_i^2M^s_j],\quad \Gamma_{11}= \sum[Q_i^2M^s_j(\bsig_i\cdot\bsig_j)] ,\\
\label{13,14}
&& \Gamma_{13}=\sum[Q_iQ_jM^s_i], \quad \Gamma_{14}=\sum[Q_iQ_jM^s_i(\bsig_i\cdot\bsig_j)],   \\
\label{19,20}
&& \Gamma_{19}=\sum[Q_i^2M^s_iM^s_j], \quad \Gamma_{20}=\sum[Q_i^2M^s_iM^s_j(\bsig_i\cdot\bsig_j)], \\
\label{25,26}
&& \Gamma_{25}=\sum[Q_iQ_jM^s_iM^s_j], \quad
 \Gamma_{26}=\sum[Q_iQ_jM^s_iM^s_j(\bsig_i\cdot\bsig_j)]. 
\end{eqnarray}
\begin{eqnarray}
&& \textrm{Three body operators:} \nonumber \\
\label{3,6,9}
&& \Gamma_3=\sum[Q_i^2(\bsig_j\cdot\bsig_k)],\  \Gamma_6=\sum[Q_iQ_j(\bsig_i+\bsig_j)\cdot\bsig_k], \  \Gamma_9=\sum[Q_i^2M^s_i(\bsig_j\cdot\bsig_k)] \\
\label{12,15,16}
&& \Gamma_{12}=\sum[Q_i^2M^s_j(\bsig_i+\bsig_j)\cdot\bsig_k], \quad \Gamma_{15}=\sum[Q_iQ_jM^s_i(\bsig_i+\bsig_j)\cdot\bsig_k], \\
\label{16,17}
&& \Gamma_{16}=\sum[Q_iQ_jM^s_k] , \quad
 \Gamma_{17}=\sum[Q_iQ_jM^s_k(\bsig_i\cdot\bsig_j)], \\
 \label{18,21}
 && \Gamma_{18}=\sum[Q_iQ_jM^s_k (\bsig_i+\bsig_j)\cdot\bsig_k], \quad \Gamma_{21}=\sum[Q_i^2M^s_iM^s_j(\bsig_i+\bsig_j)\cdot\bsig_k], \\
\label{22,23}
&& \Gamma_{22}=\sum[Q_i^2M^s_jM^s_k], \quad \Gamma_{23}=\sum[Q_i^2M^s_jM^s_k(\bsig_j\cdot\bsig_k)], \\
\label{24,27}
&& \Gamma_{24}=\sum[Q_i^2M^s_jM^s_k(\bsig_i+\bsig_j)\cdot\bsig_k], \quad
\Gamma_{27}=\sum[Q_iQ_jM^s_iM^s_j(\bsig_i+\bsig_j)\cdot\bsig_k], \\
\label{28,29}
&& \Gamma_{28}=\sum[Q_iQ_jM^s_iM^s_k], \quad \Gamma_{29}=\sum[Q_iQ_jM^s_iM^s_k(\bsig_i\cdot\bsig_k)], \\
\label{30,31}
&&\Gamma_{30}=\sum[Q_iQ_jM^s_iM^s_k(\bsig_i+\bsig_k)\cdot\bsig_j], \quad \Gamma_{31}=\sum[Q_i^2M^s_iM^s_jM^s_k], \\
\label{32}
&& \Gamma_{32}=\sum[Q_i^2M^s_iM^s_jM^s_k(\bsig_i\cdot\bsig_j + \bsig_j\cdot\bsig_k + \bsig_k\cdot\bsig_i)].
\end{eqnarray}
We have retained the original numbering of the $\Gamma$'s and have listed the structures within groups in the order of increasing numbers of factors $M^s$. There are implied unit flavor matrices and and spin operators for the quarks whose labels do not appear explicitly.  For example, $\Gamma_1$, written above as $(Q_i^2+Q_j^2+Q_k^2)$, has the complete flavor structure $(Q_i^2\openone_j\openone_k +\openone_iQ_j^2\openone_k+\openone_i\openone_jQ_k^2)$. We will indicate the presence of unit operators only where necessary for clarity. The $\Gamma$'s are independent as operators, but their matrix elements in the octet and decuplet states are not all independent: there are thirty-two operators listed but only eighteen masses.

As noted by Morpurgo \cite{Morpurgo_QM1,Morpurgo_EM2}, there are also a number of operators in which one or more of the $Q$-dependent matrix factors in the $\Gamma$s is replaced by its trace multiplied by a unit operator, for example, $Q_i^2\rightarrow\openone_i\textrm{Tr}\,Q^2$ in $\Gamma_1$. Since $\textrm{Tr}\,Q=0$, the possible trace terms are just  $\textrm{Tr}\,QM^s$, $\textrm{Tr}\,Q^2$ and $\textrm{Tr}\,Q^2M^s$.  Replacement of $M^s_i$ by $\openone_i\textrm{Tr}\,M^s=\openone_i$ simply reduces a $\Gamma$ to one with one less factor of $M^s$ and introduces nothing new. We will encounter only two of these trace terms, and do not give a listing.

Morpurgo \cite{Morpurgo_hierarchy,Morpurgo_QM2} has argued from QCD that  the coefficients of the various operators above satisfy a hierarchy of sizes, with two-body operators suppressed relative to one-body operators by the necessity of extra gluon exchanges, and three-body operators further suppressed, with extra suppressions at each stage from each symmetry breaking factor $M^s$. The results are consistent with the observed accuracy of various sum rules for the masses, including the very accurate Coleman-Glashow relation \cite{Coleman-Glashow}. Jenkins and Lebed \cite{Jenkins4,Jenkins5} have investigated the mass hierarchy in the $1/N_c$ expansion of QCD with similar but more specific results. 

Our objective here is to determine which of the $\Gamma$'s actually appear in low-order dynamical calculations, and to calculate their coefficients including the leading mesonic corrections using heavy-baryon chiral perturbation theory.  


\subsection{\label{subsec:HBChPT} Heavy-baryon effective field theory}

Our analysis of electromagnetic contributions to baryon masses will be based on heavy-baryon effective field theory with chiral meson-baryon couplings. In the heavy-baryon approximation \cite{Georgi_HBPT,Jenkins1,%
Jenkins2,Jenkins3}, the internal momenta important in a process are supposed to be small on the scale of the baryon mass, $k\ll m_B$, and baryon recoil can be neglected.  It is useful in that limit to write the momentum of a baryon in terms of its four velocity, $p^\mu=m_Bv^\mu$, $v^\mu v_\mu=1$ and to replace  the effective spin-1/2 octet and spin-3/2 decuplet baryon fields 
$B^{\gamma}$, $T^{\mu\gamma}$ in the initial chiral Lagrangian by velocity-dependent fields $B^{\,\gamma}_v$, $T^{\,\mu\gamma}_v$ defined as \cite{Georgi_HBPT,Jenkins1}
\begin{eqnarray}
\label{B_v}
B_v(x) &=& \frac{1}{2}(1+\not \!v) e^{i m_B {\not v} v^{\mu} x_{\mu}}B(x), \\
\label{T_v}
T^\mu_v(x) &=& \frac{1}{2}(1+\not \!v) e^{i m_B {\not v} v^{\mu} x_{\mu}}T^\mu(x).
\end{eqnarray}
This transformation eliminates the large momentum $m_Bv^\mu$ from the Dirac equation, and projects out particle rather than antiparticle operators. 

The velocity-dependent perturbation expansion for the redefined theory involves modified Feynman rules and an expansion in powers of $k/m_B$ \cite{Georgi_HBPT,Jenkins1}. The large mass $m_B$ does not appear directly in the new  description and there are no baryon-antibaryon vertices at leading order in $k/m_B$. As a result, a baryon always moves through a diagram in the positive time direction with its four velocity constant up to corrections of order $\bk/m_B$ in any low-momentum process. 

It will be convenient here, where we deal only with one-baryon operators, to work in the baryon rest frame with $v^\mu=(1,\,\mbox{\boldmath$0$})$.    It then becomes simple and illuminating to treat the perturbation expansion using old-fashioned time-dependent perturbation theory. We will henceforth drop the velocity labels on $B^{\,\gamma}_v$ and $T^{\,\mu\gamma}_v$ and deal only with the heavy baryon approximation at leading order.

In several earlier papers \cite{DHJ1,DHJ2, DH_moments3}, we analyzed the structure of the baryon mass and magnetic
moment operators in heavy-baryon perturbation theory (HBPT).  The analysis was greatly simplified by using  a
three-flavor-index labeling of the effective baryon fields $B_{ijk}^\gamma(x)$ and $T_{ijk}^{\mu\gamma}$,
where $i,j,k\in u,d,s$ are flavor indices and $\gamma$ is a Dirac spinor index. The transformation
properties of these fields are the same as those of the spin-1/2 and spin-3/2 operators
\begin{eqnarray}
\label{Bijk}
B_{ijk}^{\,\gamma} &\longleftrightarrow& \frac{1}{6}\,\epsilon_{abc}\,q_i^{\,\alpha a}q_j^{\,\beta\, b}q_k^{\,\gamma\, c}\,(C\gamma^5)_{\alpha \beta}\,,\\
T^{\,\mu\gamma}_{ijk}&\longleftrightarrow&\frac{1}{18\!\!\sqrt{2}}\epsilon_{abc}\left(q_i^{\,\alpha a} q_j^{\,\beta b} q_k^{\,\gamma c} + q_k^{\,\alpha a} q_j^{\,\beta b} q_i^{\,\gamma c} \right. \nonumber \\
\label{Tijk}
&& \left. + q_i^{\,\alpha a} q_k^{\,\beta b} q_j^{\,\gamma c}\right) \left(C\gamma^\mu\right)_{\alpha\beta},
\end{eqnarray}
constructed from three anticommuting ``quark'' fields  $q_i^{\,\alpha a}$, where $a\in 1,\,2,\,3$ a color index. Color will not play a role here, so we will suppress the color sums and treat the $q$'s as commuting rather than anticommuting fields. The quarks are to be treated as particles that move with the baryon with the baryon four velocity $v^\mu$. In particular, the expressions in Eqs.\ (\ref{Bijk}) and (\ref{Tijk}) reduce in the baryon rest frame to
\begin{eqnarray}
\label{Bijk'}
B_{ijk}^{\,\gamma} &\longleftrightarrow&- \frac{1}{\sqrt{6}}\left(q_i^Ti\sigma_2q_j\right)q_k^\gamma, \\
\bm{T}^{\,\gamma}_{ijk}&\longleftrightarrow&\frac{1}{6\sqrt{3}}\left[\left(q_i^T i\sigma_2\bsig q_j\right) +
\left(q_j^T i\sigma_2\bsig q_k\right)q_i^\gamma \right. \nonumber \\
\label{Tijk'}
&& \left. +\left(q_k^T i\sigma_2\bsig q_i\right)q_j^\gamma \right],
\end{eqnarray}
that is, to quark combinations with the spin-flavor structure of the SU(6) wave functions of the quark model. 

The effective octet pseudoscalar meson fields $\phi_{ij}$ correspond in this quark picture to quark-antiquark pairs in a singlet spin configuration \footnote{Note that we will not include the flavor-singlet pseudoscalar, nominally the $\eta'$, in our calculations, hence the requirement that the trace of the matrix $\phi_{ij}$ vanish.},
\begin{equation}
\label{phi_ij}
\phi_{ij} \longleftrightarrow \frac{1}{\!\!\sqrt{6}}\,\left(q_i^{\,\alpha a}\bar{q}_j^{\, \beta b} - \frac{1}{3}\delta_{ij}\, q_k^{\,\alpha a}\bar{q}_k^{\, \beta b}\right)\,\delta_{ab}\,(C\gamma^5)_{\alpha \beta}, 
\end{equation}
or, in terms of the mass eigenstates   $\phi^l$, $l \in \pi,\,K,\,\eta$, to 
\begin{eqnarray}
\label{phi_l}
 \phi^{\,l}& &= \sum_{ij}\,\lambda_{ji}^l\phi_{ij} \longleftrightarrow \frac{1}{\!\!\sqrt{6}}\,\sum_{ij}\,\lambda_{ji}^lq_i^{\,\alpha a}\bar{q}_j^{\, \beta b}\,\delta_{ab}\,(C\gamma^5)_{\alpha \beta}, \\
 \label{phi_ij2}
 \phi_{ij} &=& \frac{1}{2}\sum_l\lambda^l_{ij}\phi^l 
= \frac{1}{\sqrt{2}}\left(\begin{array}{ccc}
 \pi^0/\sqrt{2} + \eta^0/\sqrt{6}, & \pi^+, & K^+ \\
 \pi^-, & -\pi^0/\sqrt{2}+\eta^0/\sqrt{6},  & K^0 \\
 K^-, & \bar{K}^0, &-2\eta^0/\sqrt{6}
 \end{array}
 \right)
\end{eqnarray}

As we showed in \cite{DHJ1,DHJ2}, general matrix elements in heavy-baryon chiral perturbation theory (HBChPT) can be formulated using this quark picture of the baryons and mesons, and then transformed to obtain representations in terms of the elementary effective fields $B_{ijk}$, $T_{ijk}$, and $\phi^{\,l}$.  

As an example, the chiral interactions of the baryons and mesons correspond in the absence of symmetry breaking to quark-level interactions with a Lagrangian
\begin{equation}
\label{L0_quark}
\mathcal{L}_0 = i \bar{q}_iv\cdot(\mathcal{D}q)_i +  \bar{q}_i(2S^\mu A_\mu q)_i +\frac{1}{4}f^2\partial_\mu\Sigma_{ji}\partial^\mu\Sigma_{ij}^\dagger,
\end{equation}
where repeated indices are summed. Here $\mathcal{D}$ is the covariant derivative
\begin{equation}
\label{D^mu}
(\mathcal{D}_\mu q)_i = \partial_\mu q_i+\left(V_{\mu}\right)_{ii'}q_{i'},
\end{equation}
$S^\mu$ is the quark spin operator \cite{Georgi_HBPT,Jenkins1}
\begin{equation}
\label{spin_op}
S^\mu = \frac{1}{8}\left(1+\not\!v\right)\gamma^\mu\gamma^5\left(1+\not\!v\right),
\end{equation}
and $\xi$ and $\Sigma$ are flavor matrices dependent on the meson fields $\phi$,
\begin{equation}
\label{xi}
\xi=e^{i\phi/f},\quad \Sigma=e^{2i\phi/f} =\xi^2,
\end{equation}
where $f \approx 93$ MeV is the pion decay constant. $V^\mu$ and $A^\mu$ are the vector and axial vector meson current matrices
\begin{eqnarray}
\label{V^mu}
V_\mu &=& \frac{1}{2}\left(\xi\partial_\mu\xi^\dagger+\xi^\dagger\partial_\mu\xi\right) \\
&=& (1/2f^2)\left[\phi\partial_\mu\phi-(\partial_\mu\phi)\phi\right]+\textrm{O}(\phi^4/f^4), \nonumber \\
\label{A^mu}
A_\mu &=& \frac{i}{2}\left(\xi\partial_\mu\xi^\dagger-\xi^\dagger\partial_\mu\xi\right) =  f^{-1}\partial_\mu\phi + \textrm{O}(\phi^3/f^3).
\end{eqnarray}
The baryon-level Lagrangian then follows from a calculation using the definitions in Eqs.\ (\ref{Bijk}) and (\ref{Tijk}), the quark-level Lagrangian in Eq.\ (\ref{L0_quark}), and appropriate projection operators  \cite{DHJ1},
\begin{eqnarray}
\mathcal{L}_0^B &=&  i\,\bar{B}_{kji}\left( v\!\cdot\!{\cal D}\, B\right)_{ijk} + 2 \left[\bar{B}_{k'ji}\left(S^\mu\! A_\mu\right)_{k'k} B_{ijk} \right. \nonumber \\
&& \left. + \bar{B}_{kj'i}\left(S^\mu\! A_\mu \right)_{j'j}B_{ijk} \right. \nonumber \\
\label{L0_baryon}
 && \left. + \bar{B}_{kji'}\left(S^\mu\! A_\mu\right)_{i'i}B_{ijk}\right] + \ldots, 
\end{eqnarray}
where 
\begin{equation}
\label{D^mu_baryon}
{\cal D}^\nu B^{\,\gamma}_{ijk} = \partial^\nu B_{ijk}^{\,\gamma} + V_{ii'}^\nu B_{i'jk}^{\,\gamma} + V_{jj'}^\nu B_{ij'k}^{\,\gamma} + V_{kk'}^\nu B_{ijk'}^{\,\gamma}.
\end{equation}

In the remainder of this paper, we will use the quark-level description of the interactions and matrix elements, and will specialize to the baryon rest frame where  $v^\mu=(1,\bm{0})$. The quark fields reduce in that frame to two-component spinors with the spin operator  $2S^\mu=(0,\bsig)$,  and the quark Lagrangian becomes
\begin{eqnarray}
\mathcal{L}_0 &=& i\bar{q}_i\partial_0 q_i  
+ \frac{i}{2}\bar{q}_{i'}\left(\xi\partial_0\xi^\dagger+\xi^\dagger\partial_0\xi\right) _{i'i}q_{i} \nonumber \\ 
&& + \frac{i}{2}\bar{q}_{i'}\bsig\cdot\left(\xi\nabla\xi^\dagger-\xi^\dagger\nabla \xi\right)_{i'i}q_{i} \nonumber \\
&=&  i\bar{q}_i\partial_0 q_i + (1/f)\bar{q}_{i'}\bsig\cdot\nabla\phi_{i'i}q_i \nonumber \\
\label{L0_rest}
&& + (1/f^2)\bar{q}_{i'}(\phi\partial_0\phi)_{i'i}q_i + \ldots.
\end{eqnarray}
The results are general, and the expressions we encounter can be converted at any stage to covariant expressions in terms of the effective fields $B_{ijk}$ and $T^{\,\mu}_{ijk}$ in HBPT using the methods described in detail in \cite{DHJ1,DHJ2}.
 
 The $q$'s that appear in this description are not, of course, the dynamical quarks of QCD but should be thought of as structure quarks. With this interpretation, there is an additional simple connection between the effective field theory  and semirelativistic dynamical quark models for the baryons which is also useful in estimating matrix elements  \cite{DHJ1,DHJ2,DH_moments3,Morpurgo_QM2}. 

Since we will be dealing only with three-quark states, it will be convenient to suppress the quark fields and write the interaction terms in operator form. Thus, for three quarks labelled $i,\,j,\,k,$ we will write  the quark-quark-meson interaction term in Eq.\ (\ref{L0_rest}) as
 \begin{eqnarray}
 \mathcal{L}_{qqM} &=& (1/f)\left(\,\bsig_i\cdot\nabla\phi_i+\bsig_j\cdot\nabla\phi_j + \bsig_k\cdot\nabla\phi_k\,\right) \nonumber \\
 \label{Lint}
 && + (1/f^2)\left[(\phi\partial_0\phi)_i+(\phi\partial_0\phi)_j \right. \\
 && \left. +(\phi\partial_0\phi)_k\,\right] + \dots, \nonumber
 \end{eqnarray}
 where the labels on the Pauli and flavor matrices indicate the quark on which they are to act.

 This approach was very useful in our earlier analyses of the mass splittings between baryon multiplets \cite{DHJ1,DHJ2} and the baryon magnetic moments \cite{DH_moments3}. It led to simple spin-flavor descriptions of those quantities in relativistic HBPT of the type familiar in the nonrelativistic quark model and equivalent to those found by Morpurgo using his general parametrization method for matrix elements in QCD \cite{Morpurgo_param,Morpurgo_QM3,Morpurgo_EM1}. However, the connection of our results to HBPT also allowed us to calculate the dynamical one-loop mesonic contributions to the baryon masses and moments and estimate other parameters in the general expressions for those quantities in effective field theory.  We will generalize our results on the baryon masses here by including the mass splittings within the baryon and meson multiplets associated with the light quarks and the electromagnetic interactions which were previously neglected.


\subsection{\label{subsec:methods} Symmetry breaking and electromagnetic interactions}

Explicit symmetry breaking through the quark mass matrix $m=\textrm{diag}\,(m_u,m_d,m_s),$ can be incorporated by including meson and baryon mass terms proportional to  
\begin{equation}
\label{M_baryon}
\mathcal{M}=\frac{1}{2}\left(\xi^\dagger m \xi^\dagger+\xi m \xi\right)  
\end{equation}
in the chiral Lagrangian. In the case of the octet mesons, broken chiral symmetry leads to the mass term \cite{Weinberg}
\begin{equation}
\label{meson_masses1}
\mathcal{L}^M_{\textrm{mass}} = v\textrm{Tr}\,\left(\mathcal{M}-m\right) = -\frac{2v}{f^2}\textrm{Tr}\,m\phi^2 + \ldots,
\end{equation}
where $v$ is the vacuum expectation value of the quark bilinear.

The  baryon mass term in the heavy baryon limit is $\mathcal{L}_{\textrm{mass}}^B= -\bar{q}\mathcal{M}q=-\bar{q}mq+\ldots$. It is useful in this case to rewrite $m$ as  $ m_u\openone+(m_d-m_u)M^d + (m_s-m_u)M^s$ where $M^d$ and $M^s$ are the matrices $M^d=\textrm{diag}\,(0,1,0)$, $M^s=\textrm{diag}\,(0,0,1)$.   Since $m_d,\ m_u\ll m_s$, we will restrict the treatment of the the light-quark masses to first order, incorporate the term proportional to $m_u$ in the overall baryon mass parameter $m_B$, and write the single-particle mass operator at the quark level as
\begin{equation}
\label{deltam_du}
 -\Delta_{du}\left(M^d_i+M^d_j+M^d_k\right)+\tilde{\alpha}_m\left(M^s_i+M^s_j+M^s_k \right).
\end{equation}

The coefficient $\tilde{\alpha}_m\propto (m_s-m_u)$ was  introduced in the analysis of intermultiplet mass splittings in \cite{DHJ2}, $\tilde{\alpha}_m\approx 178$ MeV. The coefficient $\Delta_{du}=[(m_d-m_u)/m_s-m_u)]\tilde{\alpha}_m$ is to be determined. An evaluation of the mass ratio in terms of the meson masses \cite{Weinberg} gives
\begin{equation}
\label{d-u/s-u}
\frac{m_d-m_u}{m_s-m_u}\approx\frac{(M^2_{K^0}-M^2_{K^\pm})-(M^2_{\pi^0}-M^2_{\pi^\pm})}{M^2_{K^0}-M^2_{\pi^0}}= 0.0231,
\end{equation}
or $\Delta_{du}\approx 4.11$ MeV. 

As we will see later, there are additional purely electromagnetic contributions to the coefficient of the operator $\sum_iM^d_i$ in the complete baryon mass operator. These electromagnetic and quark-mass effects can only be untangled in fits to the mass data if the electromagnetic effects are calculable, or if $\Delta_{du}$ can be estimated as above. The coefficient of $\sum_iM^d_i$ must otherwise be treated as adjustable.

Further spin-dependent mass terms are allowed at the baryon level by general symmetry considerations, and are generated explicitly by meson loop corrections in the baryon self energy. These lead to a total baryon mass term given in operator form through one loop \cite{DHJ2} by 
\begin{eqnarray}
\label{deltaL_M}
{\cal L}^B_{\textrm{mass}} &=& - \Delta_{du}\left(\,M_i^d + M_j^d + M_k^d\,\right) \nonumber \\
&& - \tilde{\alpha}_m\left(\,M_i^s + M_j^s + M_k^s\,\right) \nonumber 
\\
&& -\frac{1}{3}\delta\tilde{m}\,
\left( \,\mbox{\boldmath$\sigma$}_i \!\cdot\! \mbox{\boldmath$\sigma$}_j + \mbox{\boldmath$\sigma$}_j\!\cdot\! \mbox{\boldmath$\sigma$}_k + \mbox{\boldmath$\sigma$}_k\!\cdot\! \mbox{\boldmath$\sigma$}_i\,\right) \nonumber 
\\
&& + \tilde{\alpha}_{ss}\left[\,(\mbox{\boldmath$\sigma$}_i+ \mbox{\boldmath$\sigma$}_j)
\!\cdot\! \mbox{\boldmath$\sigma$}_k \,M^s_k + (\mbox{\boldmath$\sigma$}_k+\mbox{\boldmath$\sigma$}_i)
\!\cdot\! \mbox{\boldmath$\sigma$}_j\,M^s_j \right. \nonumber \\
&& \left. + (\mbox{\boldmath$\sigma$}_j+\mbox{\boldmath$\sigma$}_k)
\!\cdot\! \mbox{\boldmath$\sigma$}_i\,M^s_i\,\right]  - \tilde{\alpha}_{MM}\,\left[\, M^s_iM^s_j \mbox{\boldmath$\sigma$}_{i} \!\cdot\! \mbox{\boldmath$\sigma$}_{j} \right. \nonumber 
\\
&& \left. + M^s_jM^s_k \mbox{\boldmath$\sigma$}_{j} \!\cdot\! \mbox{\boldmath$\sigma$}_{k} + M^s_kM^s_i \mbox{\boldmath$\sigma$}_{k} \!\cdot\! \mbox{\boldmath$\sigma$}_{i}\,\right].
\end{eqnarray}

To include electromagnetic effects in our analysis, we add an interaction term
\begin{eqnarray}
\mathcal{L}_{qqA} &=& -\frac{e}{2}\left(\xi Q\xi^\dagger+\xi^\dagger Q\xi\right)v^\mu A_\mu^{em} \nonumber \\
&& + \frac{e}{2}\left(\xi Q\xi^\dagger-\xi^\dagger Q\xi\right)2S^\mu A_\mu^{em}  \nonumber \\ 
&\raisebox{-1ex}{$ \stackrel{\longrightarrow}{\bm{v}\rightarrow 0}$} &-\frac{e}{2}\left(\xi Q\xi^\dagger+\xi^\dagger Q\xi\right)\Phi \nonumber \\
\label{em_int}
&&  +\frac{e}{2}\left(\xi Q\xi^\dagger-\xi^\dagger Q\xi\right)\bsig \cdot\bm{A}^{em} \\ 
&=& -e\Big(Q + (1/f^2)[\phi,[Q,\phi]]+\ldots\Big) \Phi \nonumber \\
&& +(ie/f)\Big([Q,\phi]+ \ldots\Big)\bsig  \cdot\bm{A}^{em} \nonumber
\end{eqnarray}
for each quark, and the term
\begin{eqnarray}
\label{J_meson}
\mathcal{L}_{MMA} &=& -ie\textrm{Tr}\Big(\partial^\mu\phi^\dagger[Q,\phi]+[Q,\phi^\dagger]\partial^\mu\phi\Big)A^{\textrm{em}}_\mu 
\nonumber \\
&& - e^2\textrm{Tr}[Q,\phi^\dagger][Q,\phi]A_\mu A^\mu
\end{eqnarray}
for the mesons, all in units with $\hbar=c=1$ and $\alpha_{em}=e^2/4\pi$. We will also need the quark magnetic moment interaction, given to leading order in $\phi/f$ by
\begin{equation}
\label{mom_int}
\mathcal{L}_{\textrm{mag}}=\frac{1}{2}\mu\sigma^{\lambda\nu} F_{\lambda\nu}\ \  \raisebox{-1ex}{$ \stackrel{\longrightarrow}{\bm{v}\rightarrow 0}$}\ \ \mu\bsig \cdot\nabla\times\bm{A}^{em},
\end{equation}
where $\mu=\textrm{diag}\,(\mu_u,\mu_d,\mu_s)$ is the matrix of quark magnetic moments.

\begin{figure}[ht]
\caption{\label{Fig1} Vertex diagrams for the interaction of a quark (solid line) with (a) an outgoing, and (b) an incoming meson (diagonal dashed lines), and (c) of a meson with a Coulomb field (horizontal dotted line), all at order $1/f$. Time runs upward. The vertex factors are given in Eq.\ (\protect{\ref{vertices}}).} 
\includegraphics{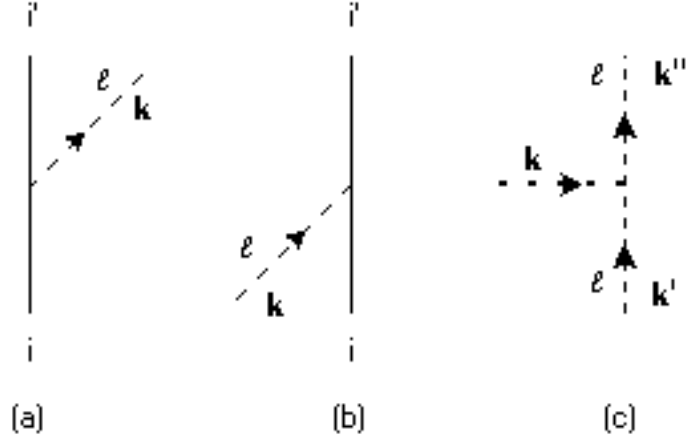}
\end{figure}

\begin{figure}
\caption{\label{Fig2} One-loop electromagnetic corrections to the baryon mass including (a) a quark self-energy contribution that can be incorporated into the unknown quark mass, (b) the Coulomb interaction between quarks, and (c), exchange of transverse photons between quarks. The diagrams in (c) vanish to $\textrm{O}(k^2/m_B^2)$ in the heavy-baryon limit when evaluated in the baryon rest frame. }
\includegraphics{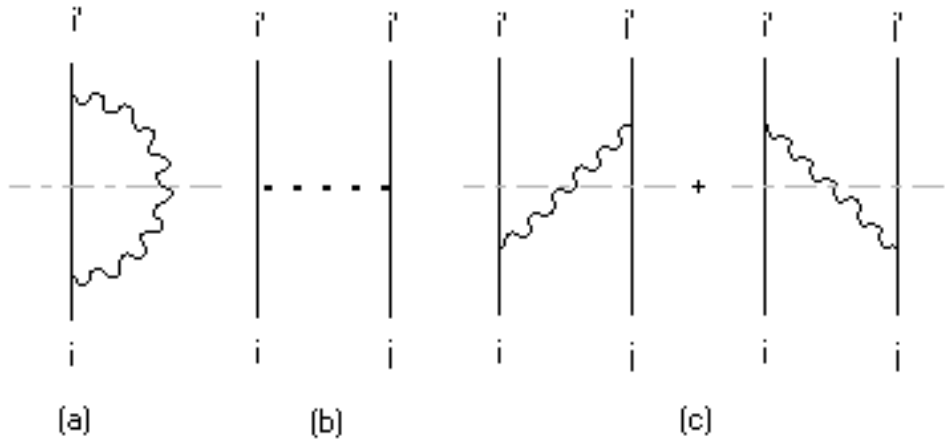}
\end{figure}

It is now straightforward to formulate the rules needed to calculate the electromagnetic contributions to the baryon masses. It is highly advantageous for our analysis to use old-fashioned time-ordered perturbation theory. This allows the clear separation of dynamically different contributions to the masses that are combined in the usual covariant approach.\footnote{The results of old fashioned perturbation theory expressed in terms integrals over the three momenta in the loops can be extracted in the covariant approach by integrating first over the time components of the loop momenta. We have used this technique in calculating the bubble diagrams in Figs.\ \ref{Fig7}\,(b) and (c) which do not fit simply into the time-ordered approach.}   We will therefore write the interactions in Hamiltonian form and will use Coulomb gauge with the  scalar potential $\Phi$ given by the instantaneous Coulomb   interaction between charges. We represent the contributions of various processes diagrammatically in Figs.\ \ref{Fig1}-\ref{Fig9}, with a solid vertical line representing a quark moving upwards toward later times, dashed lines representing mesons, and wiggly lines representing transverse photons. A horizontal dotted line represents the instantaneous Coulomb interaction between the particles on which it terminates. Each vertex is associated with a term in the interaction Hamiltonian. Thus, in Fig.\ \ref{Fig1}, 
the interactions of a quark with an outgoing or incoming meson in (a) and (b), and the charge interaction of a meson with a Coulomb field in (c) introduce factors
\begin{eqnarray}
&& (\textrm{a}) = -\frac{i}{2f}\lambda^l_{i'i}\bsig_i\cdot\bk, \quad (\textrm{b}) = \frac{i}{2f}\lambda^l_{i'i}\bsig_i\cdot\bk, \nonumber \\ 
\label{vertices}
&& (\textrm{c}) =-\frac{e}{2}Q^l\,[E_l(\bk^{''}) + E_l(\bk')],
\end{eqnarray}
where $E_l(\bk')=\sqrt{\bk^{'2}+M_l^2}$ and $E_l(\bk^{''})$ are the energies of the mesons in (c) and $Q^l$ is the meson charge.

All vertices except the Coulomb vertices are time ordered, and the contributions from all distinct orders must be included. An energy denominator $1/(E_0-E_n)$ appears in the expression for the perturbed energy for each intermediate state $|n\rangle$ of the baryon system between successive vertices, represented in Figs.\ \ref{Fig2}\,(a) and (c) by the faint horizontal lines cutting the diagrams. In the heavy-baryon approximation, the mass of the baryon cancels out of the difference $E_0-E_n$, and the energy factor reduces simply to $-1/\sum_iE_i$, where the sum is over the energies of the lines cut in the intermediate state, but with no contribution from the quark lines.

Each Coulomb line is associated with an integration 
\begin{equation}
\label{Coulomb_line}
\int\frac{d^3k}{(2\pi)^3|\bk|^2} 
\end{equation}
in momentum space or a factor $\delta(t_1-t_2)/(4\pi|\bx_1-\bx_2|)$ in position space. A meson line for a meson $l$ introduces an integration
\begin{equation}
\label{meson_line}
\int\frac{d^3k'}{(2\pi)^32E_l(\bk')}.
\end{equation}

Three momentum is conserved at internal vertices, with a factor $(2\pi)^3\delta(\sum_i\bk_i)$ at each internal vertex. There are no integrations associated with the quark lines and momentum is not conserved at a vertex on a quark line since  the quark four velocities are fixed to the baryon four velocity which does not change in leading approximation.


\section{\label{sec:corrections} Calculation of electromagnetic corrections to baryon masses}

\subsection{\label{subsec:oneloop} One-loop electromagnetic corrections}

\begin{figure}
\caption{\label{Fig3} One-loop contributions to the baryon mass that involve insertions of the mass terms in Eq.\ (\protect{\ref{deltaL_M}}). Dots represent factors of the symmetry-breaking matrix $M^s$, and zigzag lines, factors $\bsig_i\cdot\bsig_j$. Factors of $M^s$ can be omitted altogether or included at either or both vertices of a zigzag line. These diagrams cancel exactly with renormalization diagrams.}
\includegraphics{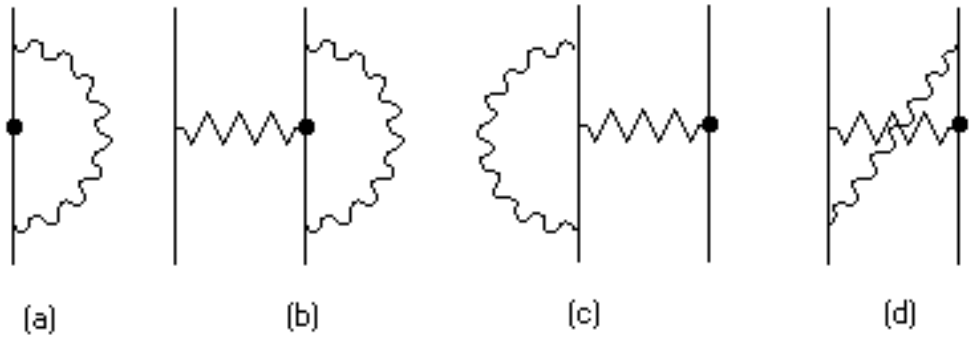}
\end{figure}

The electromagnetic corrections to the baryon masses that correspond to one-loop diagrams when viewed at the baryon level arise from the quark-level diagrams in Figs.\ \ref{Fig2} and \ref{Fig3}. The connection of the quark description to dynamical models for the baryons allows simple interpretations of the various contributions, with the diagram in Fig.\ \ref{Fig2}\,(a) corresponding to a quark self-energy diagram, 2\,(b) to the Coulomb interaction between pairs of quarks, the diagrams in Fig.\ \ref{Fig2}\,(c) to corrections arising from the emission and absorption of transverse photons between quarks, and those in Fig.\ \ref{Fig3}, to electromagnetic corrections to the initial mass terms in Eq.\ (\ref{deltaL_M}). We will use this information, and similar interpretations of other diagrams, to organize the calculations and ultimately to connect with dynamical estimates of some quantities. Recall, however, that the results are quite general and do not depend on the quark description. In particular, the matrix elements of the quark operators that appear can be transformed to covariant operator expressions in terms of  the elementary effective baryon fields of HBPT using the connection to the quark description given in Eqs.\ (\ref{Bijk'}) and (\ref{Tijk'}) and the methods developed in detail in \cite{DHJ1,DHJ2}. The results for the mass shifts are not changed.

The self-energy diagram in Fig.\ \ref{Fig2}\,(a) leads to a contribution to the baryon mass of the form $(Q_i^2 + Q_j^2 + Q_k^2)I_{\textrm{se}}$, where $I_{\textrm{se}}$ is a common integral. The charge operator is just $\Gamma_1$ in Morpurgo's parametrization, Eq.\ (\ref{1,7}). The result is equivalent to an O($e^2$) shift in the masses $m_i$ in the quark mass matrix $m=\textrm{diag}\,(m_u,m_d,m_u)$.  After a shift in the overall baryon mass $m_B$ by the up quark contribution, the result has the form of the residual quark mass term in Eq.\ (\ref{deltam_du}) and amounts to a shift in the unknown parameters $\Delta_{du}$ and $ \tilde{\alpha}_m$ in Eq.\ (\ref{deltaL_M}) and does not affect fits to the mass data. We will therefore not consider the self-energy diagram further.

The diagrams in Fig.\ \ref{Fig2}\,(c) with transverse photons involve a coupling $-eQ\bm{v}\cdot \bm{A}^{\textrm{em}}$, Eq.\ (\ref{em_int}). This vanishes in the baryon rest frame in the heavy baryon limit. More precisely, the residual coupling at each vertex is of order $|\bk|/m_B$ where $\bk$ is a typical internal momentum in the baryon, supposedly small on the scale of the baryon mass $m_B$. The diagrams are therefore of order $\bk^2/m_B^2$, and can be neglected in the heavy baryon approximation.

We are left with diagram 2\,(b) for Coulomb interactions between the quarks. This contributes a term 
\begin{equation}
\label{Coulomb_energy}
\mathcal{H}_{QQ} = e^2\left( Q_iQ_j+Q_jQ_k+Q_kQ_i\right)\int\frac{d^3k}{(2\pi)^3|\bk|^2} = I_{QQ}\Gamma_4
\end{equation}
to the effective momentum-space Hamiltonian.  Alternatively, it can be written as the position-space matrix element
\begin{equation}
\label{Coulomb_energy2}
\mathcal{H}_{QQ} = \frac{e^2}{4\pi}\left\langle\frac{Q_iQ_j}{|\bx_i-\bx_j|} + \frac{Q_jQ_k}{|\bx_j-\bx_k|} + \frac{Q_kQ_i}{|\bx_k-\bx_i|}\right\rangle.
\end{equation}
The evaluation of either expression requires further information on the structure of the baryon, or in standard HBPT, the imposition of an appropriate cutoff procedure. We will return to this problem in Sec.\ 
\ref{subsec:matrixelements}. At the moment we note only that the Coulomb energy involves the structure $\Gamma_4$ in Morpurgo's parametrization, Eq.\ (\ref{2,4,5}).

The diagrams in Fig.\ \ref{Fig3} are associated with potential electromagnetic corrections to the mass terms in Eq.\ (\ref{deltaL_M}). In these diagrams, a heavy dot indicates an insertion of a matrix $M^s$, and a zigzag line connecting quark lines $i$ and $j$, the insertion of a spin factor $\bsig_i\cdot\bsig_j$. The is no energy denominator associated with the zigzag line. Factors of $M^s$ can be inserted at neither, one, or both ends of a zigzag line leading to the three spin-dependent structures given analytically in Eq.\ (\ref{deltaL_M}).

The contributions of the transverse photons in these diagrams are of order $|\bk|^2/m_B^2$ in the baryon rest frame so would be unimportant in the heavy baryon limit. More interestingly, the diagrams cancel exactly with terms of the same order in which the mass operators have been multiplied by the wave function renormalization constant $Z=1-\delta Z$ with $\delta Z$ calculated for the photon loop, a result independent of our choice of Coulomb gauge. In particular, the photon charge matrices $Q$ commute with $M^s$ and involve no spin dependence so act as external factors with respect to the underlying mass operator. The common energy denominator $1/E_\gamma^2$ for diagrams \ref{Fig3}\,(a)-(d) is just that which appears in the $\delta Z$'s, and the product of the $Q$'s with the momentum integral reproduces the renormalization terms for a photon loop on a single quark line, and for the time-ordered photon exchange graphs. Similar cancellations will occur whenever we can slide vertices past each other to produce a diagram with the topology of an operator multiplied by a renormalization constant.


\subsection{\label{subsec:meson_exchange} Two-body corrections with meson exchange}

\begin{figure}[h]
\caption{\label{Fig4} Two-loop corrections to electromagnetic interactions that involve meson exchange between quarks. These diagrams all have the flavor structure $[Q,\lambda^l]_i[Q,\lambda^l]_j$. We do not show diagrams with transverse photons, all of which vanish in the baryon rest frame.We also do not show the diagrams of type (c) obtained by crossing one meson leg between the initial and the final state on a given quark line. These diagrams, which involve meson pair creation or absorption at the photon vertex, cancel when combined. }
\includegraphics{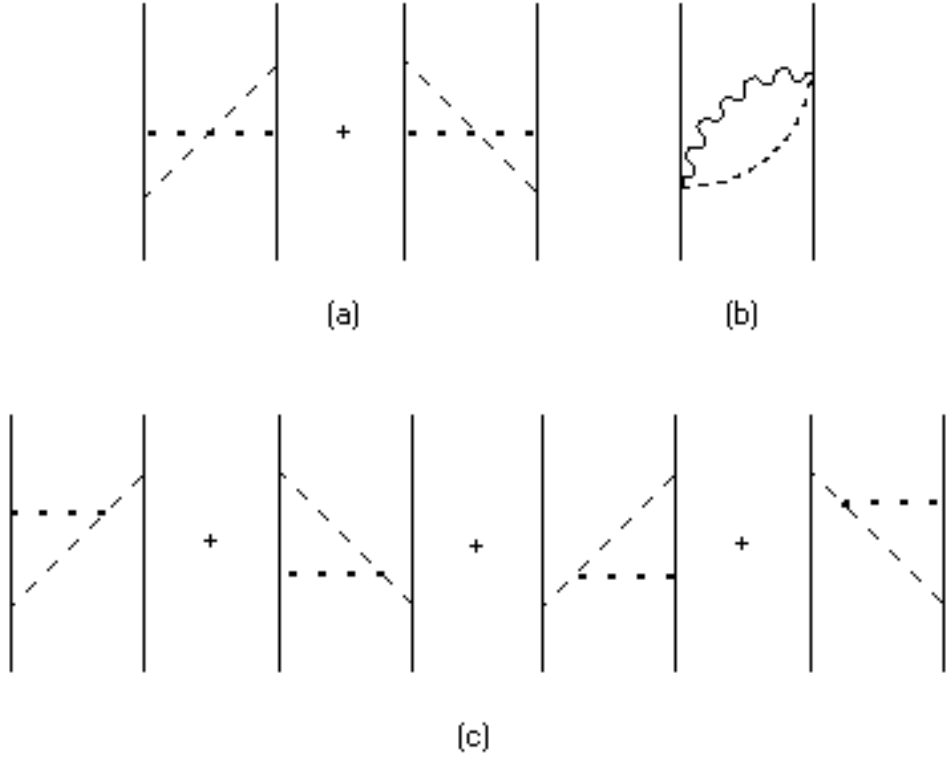}
\end{figure}

We turn next to two-body diagrams which involve the exchange of a meson between two quarks. These are shown in Fig.\ \ref{Fig4}. These are all two-loop diagrams when viewed at the baryon level, with one quark loop and one photon loop.

Figure \ref{Fig4}\,(a) shows the two time-ordered diagrams for a Coulomb interaction between quarks accompanied by a meson exchange between the same quarks. We do not show the many time-ordered diagrams that involve exchanges of a transverse photon and a meson. These vanish in the baryon rest frame in the heavy baryon limit.

Using the rules given earlier, we find a contribution
\begin{eqnarray}
&& \frac{e^2}{4f^2}\left(\lambda^l_iQ_iQ_j\lambda^l_j+ Q_i\lambda^l_i\lambda^l_jQ_j\right)
\int\frac{d^3k}{(2\pi)^3|\bk|^2} \nonumber \\
\label{double_exch}
&& \times \int \frac{d^3k'}{(2\pi)^32E_l^{'3}}\bsig_i\cdot\bk'\bsig_j\cdot\bk'
\end{eqnarray}
to the baryon mass from the diagrams in Fig.\ \ref{Fig4}\,(a) with meson $l$ exchanged. Here $E'_l=(\bk^{'2}+M_l^2)^{1/2}$. The wave function renormalization corrections from the single time-ordered meson exchanges \cite{DHJ2} change the normalization of the Coulomb contribution in Eq,\ (\ref{Coulomb_energy}) and add a term identical to that in Eq.\ (\ref{double_exch}) except thar the flavor factor is replaced by $-\left(Q_i\lambda^l_iQ_j\lambda^l_j + \lambda^l_iQ_i\lambda^l_jQ_j\right)$. The combination gives the double-exchange contribution to the effective Hamiltonian
\begin{equation}
\label{double_exch2}
\mathcal{H}_{1,ij} = -\frac{1}{3}\sum_l \,
[Q,\lambda^l]_i[Q,\lambda^l]_j \bsig_i\cdot\bsig_jI_{1,l},
\end{equation}
with similar results for the other quark pairs. Here $I_{1,l}$ is the integral left after we have used the angular integration on $\bk'$ to extract the spin dependence explicitly,
\begin{equation}
\label{I1l}
I_{1,l} = \frac{e^2}{4f^2}
\int\frac{d^3k}{(2\pi)^3|\bk|^2} \int \frac{d^3k'}{(2\pi)^32E'_l}\frac{k^{'2}}{E_l^{'2}}.
\end{equation}
 A much longer calculation in Feynman gauge gives the same result.
 
 The diagram in Fig.\ \ref{Fig4}\,(b) involves only one intermediate state with an energy denominator $-1/(E+E'_l)$ where $E$ is the energy of the transverse photon in the diagram, $E(\bk)=|\bk|$. The negative of the interaction in the last term in Eq.\ (\ref{em_int}) appears at each vertex. Since the vertices are on quark lines, three momentum is not conserved there  and the integration variables $\bk$ and $\bk'$ are independent. The  contribution to the effective Hamiltonian, including both time orders for the vertices, is simply
 \begin{equation}
 \label{joint_exchange}
 \mathcal{H}_{2,ij} = \sum_l \,[Q,\lambda^l]_i[Q,\lambda^l]_j\bsig_i\cdot\bsig_j I_{2,l},
 \end{equation}
 where $I_{2,l}$ is the integral
 \begin{equation}
 \label{I2l}
 I_{2,l} =  \frac{e^2}{4f^2}
\int\frac{d^3k}{(2\pi)^3|\bk|^2} \int \frac{d^3k'}{(2\pi)^32E'_l}\frac{E}{E+E'_l}.
\end{equation}

 The diagrams in Fig.\ \ref{Fig4}\,(c) involve a Coulomb interaction between one quark in a pair and a meson being exchanged between the quarks. The new features in this case are the appearance of the meson charge and of two integrations over internal meson lines, and of diagrams (not shown) in which a meson line is crossed from the initial to the final state or \textit{vice versa} corresponding to meson pair creation or absorption at the meson-photon vertex,. 
 
 The meson charge in a given diagram is most easily determined in terms of the quark charges. 
 Thus, in the first of Figs.\ \ref{Fig4}\,(c), the meson charge may be written in terms of the change in the charge of quark $j$ as $Q^l=Q_{j'} - Q_j$ giving a factor $(Q\lambda^l-\lambda^lQ)_{j'j}$ on that quark line, and an overall flavor factor $(\lambda^lQ)_{i'i}(Q\lambda^l-\lambda^lQ)_{j'j}$ corresponding to the operator $(\lambda^lQ)_i[Q,\lambda^l]_j$.
 
 We will choose the momenta so that the momentum of the forward-moving meson on the side of a diagram with two vertices is $\bk'$ and the momentum on the side with a single vertex is $\bk''$. There are initially integrations over $\bk'$ and $\bk''$ with the weights given in Eq.\ (\ref{meson_line}) and a momentum-conserving delta function at the three-particle vertex. We will choose the direction of the photon momentum $\bk$ so that $\bk''=\bk'+\bk$ in all diagrams. The contributions of the diagrams in Fig.\ \ref{Fig4}\,(c) can then be combined, and the total contribution to the baryon mass is
 \begin{eqnarray}
&& \frac{e^2}{4f^2}[Q,\lambda^l]_i[Q,\lambda^l]_j \int\frac{d^3k}{(2\pi)^3|\bk|^2} \int \frac{d^3k'}{(2\pi)^3} \frac{1}{4E'E''} \nonumber \\
 \label{exch_vertex}
&& \times\frac{E'+E''}{E'E''} \left(\bsig_i\cdot\bk'\bsig_j\cdot\bk''+\bsig_i\cdot\bk''\bsig_j\cdot\bk'\right)
 \end{eqnarray}
where $E'_l=E_l(\bk')$ and $E^{''}_l=E_l(\bk'')$.

This can be shown to be equivalent after angular integrations to the operator
\begin{equation}
\label{exch_vertex2}
\mathcal{H}_{3,ij} = \frac{2}{3}\sum_l\,[Q,\lambda^l]_i[Q,\lambda^l]_j\bsig_i\cdot\bsig_j I_{3,l},
\end{equation}
where $I_{3,l}$ is the integral
\begin{equation}
\label{I3l}
I_{3,l} = \frac{e^2}{4f^2} \int\frac{d^3k}{(2\pi)^3|\bk|^2} \int \frac{d^3k'}{(2\pi)^3} \frac{\bk'\cdot\bk''}{4E'E''}\frac{E'+E''}{E'E''}.
\end{equation}

Finally, the extra diagrams with creation or annihilation of meson pairs obtained from those in Fig.\ \ref{Fig4}\,(c) by crossing the meson line on the single-vertex side of the diagram from the initial to the final state, or conversely, can be shown to cancel.


\subsection{\label{subsec:meson_mass_effects} Mass and electromagnetic corrections to meson exchange}

\begin{figure}[h]
\caption{\label{Fig5} (a): The basic meson exchange diagram. (b), (c):  Electromagnetic contributions to the meson mass terms. (c): electromagnetic correction to the  meson-quark vertex. }
\includegraphics{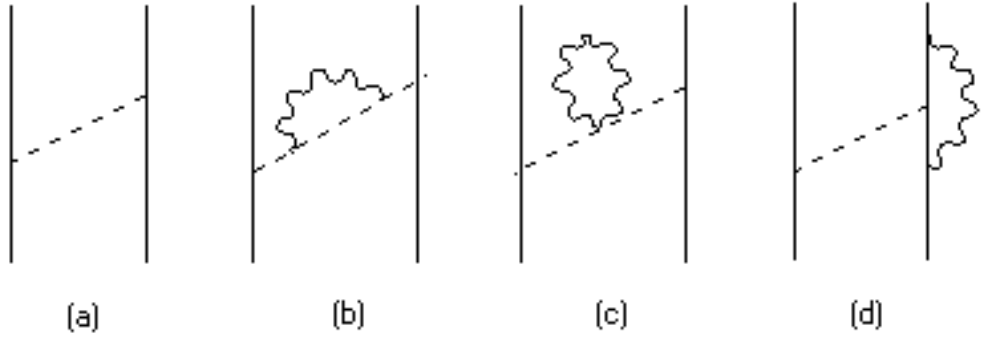}
\end{figure}

We turn next to the diagrams in Figs.\ \ref{Fig5}. Fig.\ \ref{Fig5}\,(a) depicts the 
one-loop contribution to the baryon mass from meson exchange. This was treated in  \cite{DHJ2} ignoring the mass differences within the pion and kaon multiplets and giving those multiplets appropriate average masses. The resulting baryon mass term is
\begin{equation}
\label{meson_exchange}
-\frac{1}{3}\bsig_i\cdot\bsig_j\sum_l\lambda^l_{i'i}\lambda^l_{j'j}I_l,
\end{equation}
where
\begin{equation}
\label{I_l}
I_l =  \frac{1}{4f^2}\int\frac{d^3k'}{(2\pi)^3}\frac{k^{'2}}{2E_l^{'2}} = \frac{1}{16\pi^2f^2}\int_0^\infty dk' \frac{k^{'4}}{k^{'2}+M_l^2}.
\end{equation}

This operator contributes to the splittings between the baryon isospin multiplets, but not to splittings within the multiplets. However, such intramultiplet splittings are generated by  the differences between the $\pi^\pm$ and $\pi^0$ and the $K^\pm$ and $K^0,\,\bar{K}^0$ masses. These are associated partly with the $u,\,d$ quark mass differences and partly with electromagnetic effects \cite{Weinberg}.

 From Eq.\ (\ref{meson_exchange}) or (\ref{I_l}), a change $\delta M_l^2$ in the square of the mass of meson $l$ leads to a contribution
\begin{equation}
\label{meson_mass1}
\mathcal{H}_{4,ij} =  \frac{1}{3} \bsig_i\cdot\bsig_j\sum_l\delta M_l^2\, \lambda^l_i\lambda^l_jI_{4,l},
\end{equation}
to the baryon mass operator, with $I_{4,l}$ the integral 
\begin{equation}
\label{I4l}
I_{4,l} =   \frac{1}{4f^2}\int \frac{d^3k'}{(2\pi)^32E'} \frac{k^{'2}}{E'^3}.
\end{equation}

The mass shifts $\delta M_l^2$ are known \cite{Weinberg}. The quark-mass contributions to the meson masses are given by Eq.\ (\ref{meson_masses1}) with $M_l^2=(v/2f^2) \textrm{Tr}\,m\lambda^l\lambda^l$ giving
\begin{eqnarray}
 M^2_{\pi^\pm} &=& M^2_{\pi^0}=\frac{v}{2f^2}(m_u+m_d), \nonumber \\
 M_{K^\pm}^2 &=& \frac{v}{2f^2}(m_u+m_s),\quad M_{K^0}^2=\frac{v}{2f^2}(m_d+m_s), \nonumber \\
 \label{meson_masses2}
 M_{\eta^0}^2 &=&  \frac{v}{6f^2}(m_u+m_d+4m_s),
 \end{eqnarray}
 The only intramultiplet splitting associated with the quark masses is therefore in the kaon system where
 \begin{equation}
 \label{Ksplitting1}
 M_{K^\pm}^2-M_{K^0}^2 = -\frac{v}{2f^2}(m_d-m_u)\equiv -\Delta^ M_q.
 \end{equation}

 We will choose $M_{K^0}$ as the mass $M_l$ for the kaon exchange diagrams and $\delta M_K^2$ as $M_{K^\pm}^2-M_{K^0}^2$. The contribution of quark masses to the total mass splitting $\delta M_K^2$ is then given by the right hand side of Eq.\ (\ref{Ksplitting1}). The correction exists only for $K^\pm$ exchange. We can isolate this from the general expression in Eq.\ (\ref{meson_mass1}) by specifying the initial and final quarks at a vertex using the matrices $M^u$ and $M^s$ as projection operators and tracing the flavors through the diagram. Thus, for the time ordered exchange diagram in Fig.\ \ref{Fig5}\,(a), $K^+$ exchange requires that $i=j'=u,\,j=i'=s$ corresponding to 
 a flavor factor $2M^u_{j'i}M^s_{i'j}$.  For the same time ordering of the vertices, $K^-$ exchange corresponds  to a flavor factor $2M^s_{j'i}M^u_{i'j}$. The total contribution to $\mathcal{H}_{4,ij}$ including $K^+$ and $K^-$ exchange with both time orders is
 \begin{equation}
 \label{meson_mass2}
 -\frac{4}{3} \Delta^ M_q I_{4,K^0}\,\left(M^u_{j'i}M^s_{i'j}+M^s_{j'i}M^u_{i'j}\right)  \bsig_i\cdot\bsig_j.
\end{equation}

The electromagnetic interactions in Eq.\ (\ref{J_meson}) also contribute to the meson mass differences through loop diagrams. In particular, the diagrams in Figs.\ \ref{Fig5}\,(b) and (c) represent electromagnetic corrections to the meson mass operators which, when evaluated with the meson on-shell, give contributions to the meson mass. We will ignore the dependence of the loops on the incoming momentum $\bk$ for simplicity, and will treat these diagrams in terms of electromagnetic contributions to the physical meson masses.   

The expression for the charge of a meson $\phi^l$ involves a commutator $[Q,\phi^l]$ which vanishes for the diagonal $\pi^0$ and $\eta^0$ terms in the $\phi$ matrix, Eq.\ (\ref{phi_ij2}), and also for the $K^0,\,\bar{K}^0$ terms because of the equality of the $d$ and $s$ charges. The contributions from Figs.\ \ref{Fig5}\,(b) and (c) are proportional to $[Q,\lambda^l]_i[Q,\lambda^l]_j$ so vanish for the $\pi^0$,  $\eta^0$, $K^0$,  and $\bar{K}^0$. The charges in the $\pi^\pm$ and $K^\pm$ systems are the same, so the loop corrections for those particles are the same up to corrections involving the different meson masses in the loops. We will not attempt to calculate the electromagnetic corrections, but will simply denote them by $\Delta^M_{\textrm{em}}$, ignoring the presumably small difference for the pions and kaons. This gives
\begin{equation}
\label{meson_mass3}
M_{\pi^\pm}^2-M_{\pi^0}^2=\Delta^M_{\textrm{em}},\quad  M_{K^\pm}^2-M_{K^0}^2 =  - \Delta^M_q+ \Delta^M_{\textrm{em}}.
\end{equation}

Choosing $M_{\pi^0}$ as the mass $M_l$ in the pion exchange diagrams and constructing the flavor factor as above, we find that the contribution to baryon mass splittings from the meson exchange diagrams is
\begin{eqnarray}
\mathcal{H}_{4,ij} &=&  \frac{4}{3} \Delta^M_{\textrm{em}}I_{4,\pi^0}\,\left(M^u_{j'i}M^d_{i'j} + M^d_{j'i}M^u_{i'j}\right)  \bsig_i\cdot\bsig_j \nonumber \\
&& +\frac{4}{3}(\Delta^M_{\textrm{em}}- \Delta^M_q) I_{4,K^0}\ \nonumber \\
\label{meson_mass4}
&& \times\left(M^u_{j'i}M^s_{i'j}+M^s_{j'i}M^u_{i'j}\right)  \bsig_i\cdot\bsig_j.
\end{eqnarray}

While this form for the electromagnetic part of the correction is simple, it does not display its electromagnetic character. We will therefore use the alternative expression 
\begin{eqnarray}
\mathcal{H}_{4,ij} &=& - \frac{4}{3}\Delta^M_q I_{4,K^0}\,\left(M^u_{j'i}M^s_{i'j}+M^s_{j'i}M^u_{i'j}\right)  \bsig_i\cdot\bsig_j \nonumber \\
\label{meson_mass5}
&& -\frac{2}{3}\Delta^M_{\textrm{em}}\sum_l[Q,\lambda^l]_{i'i}[Q,\lambda^l]_{j'j} I_{4,l}\,\bsig_i\cdot\bsig_j
\end{eqnarray}
that displays the charges explicitly and connects directly to Morpurgo's general parametrization of the electromagnetic contributions to the baryon masses. The equivalence of the expressions may be seen when the reduced form of the electromagnetic term given later in Eq.\ (\ref{reduc[q,lambda]^2}) is evaluated using the actual quark charges.

\begin{figure}
\caption{\label{Fig6} A baryon $Z$ graph corresponding to the vertex correction in Fig.~\protect{ \ref{Fig5}}\,(d) with an instantaneous Coulomb interaction rather than a transverse photon encompassing the meson-quark vertex.}
\includegraphics{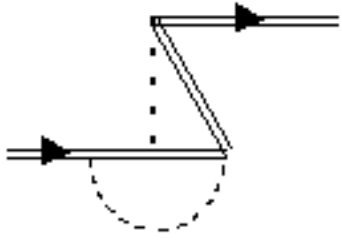}
\end{figure}

Finally, there is an electromagnetic vertex correction to the meson exchange diagram as shown in Fig.\ \ref{Fig5}\,(d). This combines with the corresponding renormalization diagrams to give a contribution with a flavor factor $\lambda_i[Q,[\lambda^l,Q]]_j$. This structure is actually not new because of the identity
\begin{equation}
\label{QlambdaQ_iden}
\lambda_i[Q^l,[\lambda^l,Q]]_j =  [Q,\lambda^l]_i [Q,\lambda^l]_j,
\end{equation}
a result that may be derived in the context of Fig.\ \ref{Fig4}\,(c) by evaluating the meson charge alternately in terms of the quark charges on the line with two vertices and on the line with a single vertex. It may be proved directly  by using the techniques to be discussed in \S \ref{subsec:reduction} to eliminate the Gell-Mann matrices on the two sides of the equation. The results are identical. While this would allow us to combine the contribution from  Fig.\ \ref{Fig5}\,(b) with those from Fig.\,\ref{Fig4}  as far as analyzing its structure is concerned, this contribution actually vanishes in the heavy baryon limit. The vertices that involve transverse photons give factors $\bv\cdot\bA$, so vanish in the baryon rest frame for constant baryon or quark four velocity $v^\mu=(v^0,\bv)\rightarrow(1,\bm{0})$. Instantaneous Coulomb interactions enter only through ``Z graphs'' such as that shown on the baryon level in Fig.\ \ref{Fig6}. These involve the creation or annihilation of baryon pairs and are suppressed by an extra  term $2m_B$ in an energy denominator and vanish for $m_B\rightarrow\infty$. 
%


\subsection{\label{subsec:vertexcorrections} Meson vertex corrections}

\begin{figure}[h]
\caption{\label{Fig7} Mesonic corrections to electromagnetic vertices. All diagrams have the $Q_i[\lambda^l,[Q,\lambda^l]]_j$ flavor structure.}
\includegraphics{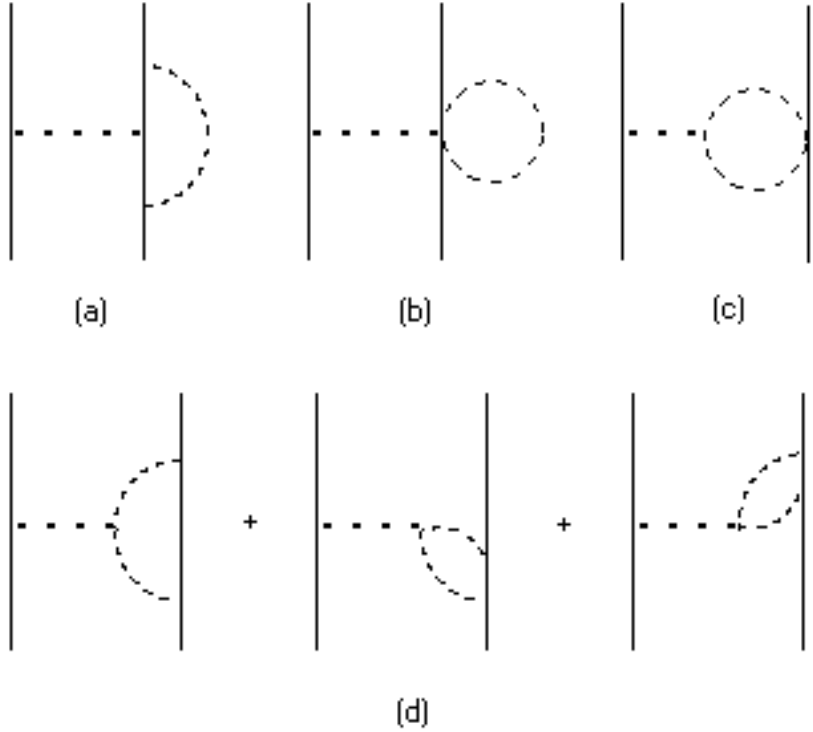}
\end{figure}

The meson loop corrections to the electromagnetic vertices are shown in Fig.\ \ref{Fig7}. All are proportional to 
the flavor factor $Q_i[\lambda^l,[Q,\lambda^l]]_j$. Thus, the diagram in Fig.\ \ref{Fig7}\,(a) and the corresponding renormalization diagrams give 
\begin{equation}
\label{meson_vert1}
\mathcal{H}_{5,ij} = \frac{1}{2}\sum_l Q_i[\lambda^l,[Q,\lambda^l]]_j I_{5,l},
\end{equation}
where the integral $I_{5,l}\equiv I_{1,l}$ is defined in Eq.\ (\ref{I1l}). The equality of the integrals associated with the diagrams in Figs.\ \ref{Fig4}\,(a) and \ref{Fig7}\,(a) is not surprising: both are components of a single baryon-level diagram with a Coulomb insertion on the baryon line inside a meson loop.

The diagram in Fig. \ref{Fig7}\,(b) arises from the first chiral correction to the electromagnetic current in Eq.\ (\ref{em_int}). Its contribution can be evaluated using our rules even in the absence of an energy denominator, or less directly by starting with the covariant expression for the Feynman-gauge diagrams formulated using the heavy baryon approach of Jenkins and Manohar \cite{Jenkins1,Jenkins3} and integrating over the timelike component $k^{'0}$ of the meson loop momentum. The result is
\begin{equation}
\label{chiral_loop}
\mathcal{H}_{6,ij} = \frac{1}{2}\sum_l Q_i[\lambda^l,[Q,\lambda^l]]_j I_{6,l},
\end{equation}
where
\begin{equation}
\label{I6l}
I_{6,l} = \frac{e^2}{4f^2} \int\frac{d^3k}{(2\pi)^3|\bk|^2} \int \frac{d^3k'}{(2\pi)^32E'_l}.
\end{equation}

The right-hand vertex in Fig.\ \ref{Fig7}\,(c) is associated with the vector current in Eq.\ (\ref{V^mu}) and introduces a factor $v\cdot(k'+k'')/2$ in an arbitrary Lorentz frame or $(E'+E'')/2$ in the baryon rest frame. The meson electromagnetic vertex in the center introduces a further factor  $v\cdot(k'+k'')\rightarrow(E'+E'')$. The flavor factor involves the charge of the meson and can be evaluated either as $[Q,\lambda^l]\lambda^l$ or as $-\lambda^l[Q,\lambda^l]$. The two expressions are equivalent with
\begin{equation}
\label{lambdaQlambda}
[Q,\lambda^l]\lambda^l=-\lambda^l[Q,\lambda^l]=-\frac{1}{2}[\lambda^l,[Q,\lambda^l]].
\end{equation}
We will use the double commutator form in writing the matrix factors. The loop integrals can be calculated in the covariant approach as sketched above for Fig.\ \ref{Fig7}\,(b) with the result 
\begin{equation}
\label{V_bubble}
\mathcal{H}_{7,ij} =  -\frac{1}{2}\sum_l Q_i[\lambda^l,[Q,\lambda^l]]_j I_{7,l}
\end{equation}
with
\begin{equation}
\label{I7l}
I_{7,l} = \frac{e^2}{4f^2} \int\frac{d^3k}{(2\pi)^3|\bk|^2} \int \frac{d^3k'}{(2\pi)^3}\frac{1}{E'_l+E''_l}.
\end{equation}

The vertex and pair creation and annihilation diagrams in Fig.\ \ref{Fig7}\,(d) are related by a crossing transformation which changes the initial electromagnetic vertex factor $E'+E''$ in the first diagram to $E'-E''$ in the second, and to $-E'+E''$ in the third. The energy denominators also change. As a result, the creation and annihilation diagrams do not cancel, in contrast to the corresponding diagrams that exist, but are not shown, for Fig.\ \ref{Fig4}\,(c). The combination of the three diagrams in Fig.\ \ref{Fig7}\,(c) gives a contribution
\begin{equation}
\label{meson_vert2}
\mathcal{H}_{8,ij} = -\frac{1}{2}\sum_l Q_i[\lambda^l,[Q,\lambda^l]]_j I_{8,l}
\end{equation}
to the effective mass Hamiltonian, with $I_{8,l}$ given by
\begin{equation}
\label{I8l}
I_{8,l} = \frac{e^2}{4f^2} \int\frac{d^3k}{(2\pi)^3|\bk|^2} \int \frac{d^3k'}{(2\pi)^3}\frac{\bk'\cdot\bk''}{E'E''} \frac{4}{E'+E''}
\end{equation}
after dropping terms that vanish in the angular integrations.

It is easily checked that the sum of the mesonic corrections to the photon-quark vertex vanishes as $\bk^2$ for $\bk\rightarrow 0$, a consequence electromagnetic current conservation, so the quark and baryon charges are not changed by the corrections.


\subsection{\label{subsec:magneticmoments} Magnetic moment interactions}

\begin{figure}[h]
\caption{\label{Fig8} Instantaneous magnetic moment-moment interactions and mesonic corrections. A zigzag line with crosses at the vertices represents a factor $\mathcal{H}_{\mu\mu}$ from Eq.~(\protect{\ref{H_mumu1}}) or (\protect{\ref{H_mumu2}}). }
\includegraphics{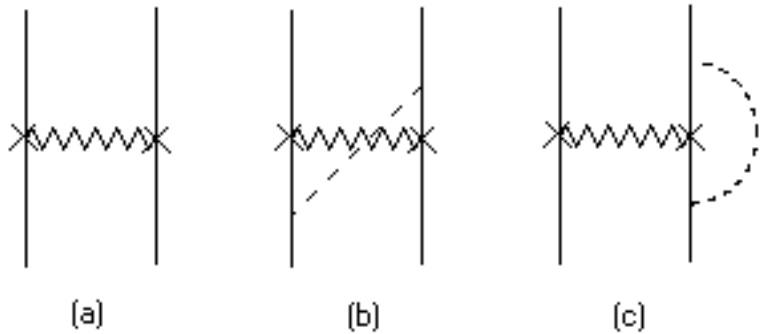}
\end{figure}

We turn finally to the direct interaction between magnetic moments depicted in Fig.\ \ref{Fig8}. 
The leading term in Fig.~\ref{Fig8}\,(a) contributes a term to the effective Hamiltonian given in momentum space by

\begin{eqnarray}
\mathcal{H}_{\mu\mu;ij} &=& -4\pi\int\frac{d^3k}{(2\pi)^3}\frac{\mu_i\mu_j(\bk\times\bsig_i)\cdot(\bk\times\bsig_j)}{|\bk|^2} \nonumber \\
\label{H_mumu1}
&=& -\frac{8\pi}{3}\mu_i\mu_j\bsig_i\cdot\bsig_j\int\frac{d^3k}{(2\pi)^3} \\
&& \equiv -\frac{8\pi}{3}\mu_i\mu_j\bsig_i\cdot\bsig_j\,I_{\mu\mu} \nonumber
\end{eqnarray}
or, in position space, by
\begin{equation}
\label{H_mumu2}
\mathcal{H}_{\mu\mu;ij} = -\frac{8\pi}{3}\mu_i\mu_j\bsig_i\cdot\bsig_j\, \delta^3(\bx_i-\bx_j).
\end{equation}
The extra factor of $4\pi$ in Eq.\ (\ref{H_mumu1}) and the following equations arises from the conversion of the moments from rationalized units to nuclear magnetons which we will use below.

We analyzed the form of the magnetic moments in effective field theory in detail in \cite{DH_moments3}, including the one-loop mesonic corrections. These rather small corrections could be included in the $\mu_i$ in the leading moment-moment interaction above, but are not numerically significant. They correspond to corrections to corrections as far as the diagrams in Fig.\ \ref{Fig8}\,(b) and (c) are concerned and will be neglected there as well. It is sufficient for all these diagrams to use the quark-model moments
\begin{equation}
\label{QM_moments}
\mu_i = \mu_aQ_i + \mu_b(QM^s)_i,
\end{equation}
where $\mu_a=2.793$ nm and $\mu_b=-0.933$ nm.

After considerable spin algebra and angular integration, the meson-exchange correction to the moment-moment interaction depicted in Fig.\ \ref{Fig8}\,(b) reduces to
\begin{eqnarray}
\label{moment_exchange}
\mathcal{H}_{9,ij} &=& \sum_l\left( [\mu,\lambda^l]_i[\mu,\lambda^l]_j \right. \nonumber \\
&& \left. - \frac{2}{3}\left(\{\mu,\lambda^l\}_i \{\mu,\lambda^l\}_j\right)\bsig_i\cdot\bsig_j\right)I_{9,l},
\end{eqnarray}
where $\mu_i$ is given above and
\begin{equation}
\label{I9l}
I_{9,l} = \frac{8\pi}{3}\delta^3(\bx_i-\bx_j)\frac{1}{4f^2}\int\frac{d^3k'}{(2\pi)^32E'}\frac{k^{'2}}{E^{'2}}.
\end{equation}

The diagram in Fig.\ \ref{Fig8}\,(c) gives a similar result,
\begin{eqnarray}
\label{moment_vertex}
\mathcal{H}_{10,ij} &=& \sum_l\left(\frac{1}{6}\mu_i[\lambda^l,[\mu,\lambda^l]]_j \right. \nonumber \\
&& \left. + \frac{2}{3}\mu_i\left(\mu\lambda^l\lambda^l+\lambda^l\lambda^l\mu\right)_j \right)\bsig_i\cdot\bsig_j I_{ 10,l},
\end{eqnarray}
where $I_{10,l}=I_{9,l}$.


\subsection{\label{subsec:threebody} Three-body diagrams}

\begin{figure}[h]
\caption{\label{Fig9} The three-body, two-loop diagrams involving electromagnetic interactions. We do not show the many diagrams with transverse photons corresponding to the Coulomb diagrams in (a) and (b) or the other possible time ordering for (c)-(d) as these diagrams all vanish in the baryon rest frame. The mass insertion (heavy dot) in (d) can be omitted or included at either or both vertices as in Fig.~\protect{\ref{Fig3}}. }
\includegraphics{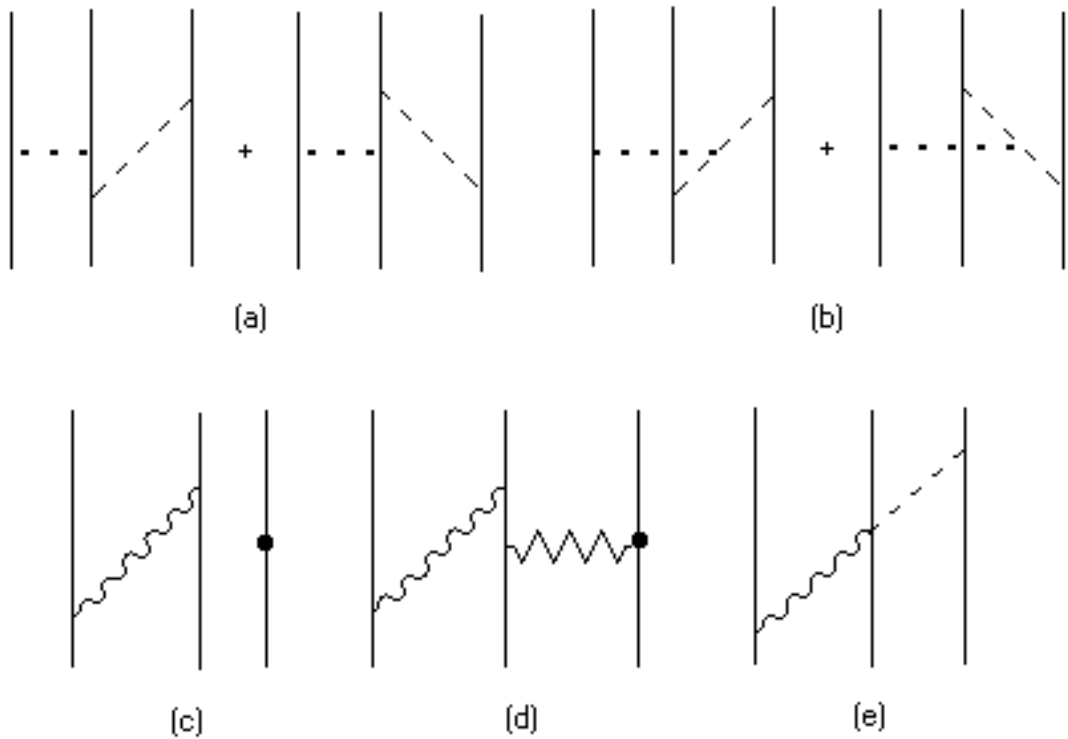}
\end{figure}

The two-loop three-body diagrams for the charge interactions  are shown in Fig.\ \ref{Fig9}. The double
exchange diagrams in Fig.\ \ref{Fig9}\,(a) are not intertwined, and it is easy to see that these diagrams cancel exactly with the renormalization diagrams with the same topology. The same is true of the diagrams with mass insertions in Figs.\ \ref{Fig9}\,(c) and (d), and of the diagrams, not shown, in which the insertion in 9(d) has the factor $M^s$ at the opposite vertex on the zigzag line or is replaced by a moment-moment interaction. 

The two diagrams in Fig.\ \ref{Fig9}\,(b) involve quark charge interactions with internal meson currents, with the two possible time orderings shown. The different orders correspond to the exchanges of mesons of opposite charge, and the contributions of the two diagrams cancel exactly.

Finally, the diagram in Fig.\ \ref{Fig9}\,(e), which arises from the last term in Eq.\ (\ref{em_int}), vanishes in the baryon rest frame in the heavy baryon limit because of a factor $\bv$ in the quark-photon coupling.

We find, then, the important result that there are \textit{no} three-body electromagnetic contributions to the baryon masses through one loop (order $1/f^2$) in the mesonic corrections. The first three-body corrections arise  at three loops overall, for example, from the addition of a meson line connecting a vertex between the existing vertices  on either of lines $i$ or $j$ in the first of Figs.\ \ref{Fig4}\,(a) to the third quark line, $k$. In general, for a three-body correction to be nonzero, it must not be possible to slide vertices in the corresponding three-loop diagram past each other to obtain a diagram with the topology of a renormalization correction to a one- or two-loop diagram.


\section{\label{sec:splittings} Intramultiplet baryon mass splittings}

\subsection{\label{subsec:reduction} Reduction to basic structures}

We wish at this point to identify the independent structures that appear in the electromagnetic corrections to the baryon masses through two loops (order $e^2/f^2$), and to identify them with the structures in Morpurgo's parametrization in Eqs.\ (\ref{sums})-(\ref{32}). We note first that there are no three body contributions to this order, so the structures in Eqs.\ (\ref{3,6,9})-(\ref{32}) are absent. The basic charge and magnetic moment interactions in Eqs.\ (\ref{Coulomb_energy}) and (\ref{H_mumu1}), with the moments in Eq.\ (\ref{QM_moments}), give the combinations
\begin{eqnarray}
\label{reduction1}
\mathcal{H}_{QQ}&=&I_{QQ}\Gamma_4,  \\
\label{reduction2}
\mathcal{H}_{\mu\mu}&=&\left(\mu_a^2\Gamma_5+2\mu_a\mu_b\Gamma_{14}+\mu_b^2\Gamma_{26} \right) I_{\mu\mu}.
\end{eqnarray}
We will consider the mesonic corrections to these structures separately.


\subsubsection{\label{subsubsec:charge} Corrections to the charge interactions}

The diagrams in Figs.\ \ref{Fig4} and \ref{Fig5}\,(a-c) all lead to contributions of the form
\begin{equation}
\label{reduction3}
\sum_l\,[Q,\lambda^l]_i[Q,\lambda^l]_j\bsig_i\cdot\bsig_jI_l.
\end{equation}
To reduce this to standard form, we use the identity derived in \cite{DHJ2}, that\footnote{The terms in $I_\pi$ and $I_K$ involving two factors of $M^s$ were rewritten in \cite{DHJ2} using the identity $M^s_{i'j}M^s_{j'i} = M^s_{i'i}M^s_{j'j}$.}
\begin{eqnarray}
\sum_l\lambda^l_{i'i}\lambda^l_{j'j}\,I_l &=& 2I_\pi\left(\openone-M^s\right)_{i'j}\left(\openone-M^s\right)_{j'i} \nonumber \\
&& - I_\pi\left(\openone-M^s\right)_{i'i}\left(\openone-M^s\right)_{j'j} \nonumber \\
&& +2I_K\left[\left(\openone-M^s\right)_{i'j}M^s_{j'i} \right. \nonumber \\
&& \left. + M^s_{i'j}\left(\openone-M^s\right)_{j'i}\right] \nonumber  \\
\label{I_lambda'_lambda}
&& +\frac{1}{3}I_\eta\left(\openone-3M^s\right)_{i'i}\left(\openone-3M^s\right)_{j'j}. 
\end{eqnarray}
We find that 
\begin{eqnarray}
\sum_l [Q,\lambda^l]_{i'i}[Q,\lambda^l]_{j'j}I_l &=& 2I_\pi\left(2Q_{i'j}Q_{j'i} 
 -\openone_{i'j}Q^2_{j'i}-Q^2_{i'j}\openone_{j'i}\right) \nonumber \\
&&-2(I_\pi-I_K) \left[2Q_{i'j}(QM^s)_{j'i}+2(QM^s)_{i'j}Q_{j'i} \right. \\
\label{reduc[q,lambda]^2}
&& \left. - Q^2_{i'j}M^s_{j'i} - M^s_{i'j}Q^2_{j'i} \right. \\ 
&& \left. -(Q^2M^s)_{i'j} \openone_{j'i}-\openone_{i'j}(Q^2M^s)_{j'i}\right].  \nonumber
\end{eqnarray}
The skew structure, with the indices connected in the pairs $i',\,j$ and $j',\,i$, can be interpreted simply in the quark picture: in the case of a charged meson exchange the incoming quark lines must run continuously through the diagrams from the initial to the final states, and must be interchanged in the final state because of the exchange. Explicit evaluation of this expression for $Q_u=2/3$, $Q_d=Q_s = -1/3$ shows that the only contributions are from the $\pi^\pm$ and $K^\pm$ as expected from the commutator structure.

Multiplying Eq.\ (\ref{reduc[q,lambda]^2}) by the factor $\bsig_i\cdot\bsig_j$ and using the exchange operator $P_{ij}=(1+\bsig_i\cdot\bsig_j)/2$ from \cite{DHJ2} to rearrange indices,
\begin{equation}
\label{exchange_op}
A_{i'j;j'i} \, \bsig_i \cdot \bsig_j  =
A_{i'i;j'j} \, P_{ij} \, \bsig_i \cdot \bsig_j =
A_{i'i;j'j} \left(\frac{3 }{2} -\frac {1}{2} \bsig_i \cdot \bsig_j\right),
\end{equation}
we find that
\begin{eqnarray}
\sum_l [Q,\lambda^l]_{i'i}[Q,\lambda^l]_{j'j}I_l\bsig_i \cdot \bsig_j&=& I_\pi\left[2Q_{i'i}Q_{j'j}-Q^2_{i'i}\openone_{j'j} - \openone_{i'i}Q^2_{j'j}\right]\left(3- \bsig_i \cdot \bsig_j\right) \nonumber \\
&& -(I_\pi-I_K)\left[2Q_{i'i}(QM^s)_{j'j}+2(QM^s)_{i'i}Q_{j'j} \right. \nonumber \\
&& \left.  -Q^2_{i'i}M^s_{j'j}-M^s_{i'i}Q^2_{j'j}\right]\left(3- \bsig_i \cdot \bsig_j\right) -(I_\pi-I_K) \nonumber \\
\label{reduc[q,lambda]^2_2}
&&\times \left[(Q^2M^s)_{i'i}\openone_{j'j} +\openone_{i'i}(Q^2M^s)_{j'j}\right]\left(3- \bsig_i \cdot \bsig_j\right) .  
\end{eqnarray}
Adding the pieces for the remaining pairs of quarks and identifying the results with the structures in Eqs.\ (\ref{1,7}-\ref{13,14}), we find that the result contains all ten of Morpurgo's one- and two-body operators $\Gamma_i$ with at most one factor of $M^s$. 

The spin-independent and spin-dependent contributions from the last row, proportional to $\Gamma_7$ and $\Gamma_8$ respectively,  do not contribute to the mass splittings within isospin multiplets. Furthermore, $Q^2M^s=(1/9)M^s$ acts only on strange quarks, so $\Gamma_7$ and $\Gamma_8$ are proportional to the operators  in the original intermultiplet mass terms in Eq.\ (\ref{deltaL_M}) with the coefficients  $\tilde{\alpha}_m$ and $\tilde{\alpha}_{ss}$. The addition of the $\Gamma_7$ and $\Gamma_8$ electromagnetic corrections is equivalent to changing $\tilde{\alpha}_m$ and $\tilde{\alpha}_{ss}$. However, those coefficients depend on short-distance interactions  which are not known, and are therefore treated as parameters in fits to the baryon mass spectrum \cite{DHJ2,DH_masses}. The addition of the electromagnetic corrections does not change the fits  after readjustment of the parameters. We will therefore drop terms in $\Gamma_7$ and $\Gamma_8$.

The spin-independent term $\sum_iQ^2_i= \Gamma_1$ from the first row of Eq.\ (\ref{reduc[q,lambda]^2_2})  can also be reduced, in this case to the form of the quark mass-difference term in Eq.\ (\ref{deltam_du}).  In particular,
\begin{eqnarray}
\Gamma_1 &=& \sum_iQ_i^2 = Q_u^2\sum_i\openone_i+ \left(Q_d^2-Q_u^2\right)\sum_iM_i^d  \nonumber \\
\label{reduction_Gamma1}
&& + \left(Q_s^2-Q_u^2\right) \sum_iM_i^s.
\end{eqnarray}
The parts of the electromagnetic corrections proportional to the unit operator can be absorbed by adjusting the overall baryon mass parameter $m_B$, while those proportional to $\sum_i M^s_i$ can be absorbed in $\tilde{\alpha}_m$. The operator $\sum_i M^d_i$ has the form of the quark mass-difference operator in Eq.\ (\ref{deltam_du}, but this part of the electromagnetic correction arises from meson exchange effects rather than quaark self energies and is significant to the extent that $\Delta_{\textrm{du}}$ can be assumed to be known from the estimate in Eq.\ (\ref{d-u/s-u}).

Upon collecting the relevant integrals, reducing $\Gamma_1$ as above, and dropping the terms in $\Gamma_7$ and $\Gamma_8$, we find that the new contributions to the baryon mass from the diagrams in Figs.\ \ref{Fig4} and \ref{Fig5} are given by the operator
\begin{eqnarray}
&& 6\mathcal{I}_{1,\pi} Q_iQ_j + \mathcal{I}_{1,\pi}(M^d_i\openone_j+\openone_iM^d_j) -\mathcal{I}_{1,\pi} \left(2Q_iQ_j-Q^2_i\openone_j-\openone_i Q^2_j\right)\bsig_i\cdot\bsig_j   \nonumber \\
&& \ \ -(\mathcal{I}_{1,\pi}-\mathcal{I}_{1,K})\left[2Q_i(QM^s)_j+2(QM^s)_iQ_j-Q^2_iM^s_j-M^s_iQ^2_j\right] \nonumber \\
&& \times (3-\bsig_i\cdot\bsig_j ) +\textrm{perms} \nonumber \\
&& = 6\mathcal{I}_{1,\pi}\Gamma_4 + 2\mathcal{I}_{1,\pi}(M^d_i+M^d_j+M^d_k) -2\mathcal{I}_{1,\pi}\left(\Gamma_5-\Gamma_2\right) \nonumber \\ 
\label{reduc[q,lambda]^2_3}
&& \ \ -2(\mathcal{I}_{1,\pi}-\mathcal{I}_{1,K})\left(6\Gamma_{13}-3\Gamma_{10}-2\Gamma_{14}+\Gamma_{11}\right),
\end{eqnarray}
where 
\begin{equation}
\label{I_(4,5)}
\mathcal{I}_{1,l }= -\frac{1}{3}I_{1,l}+I_{2.l}+\frac{2}{3}I_{3,l}-\frac{2}{3}\Delta^M_{\textrm{em}}I_{4,l}.
\end{equation}

We can reduce the contributions from the diagrams in Fig.\ \ref{Fig7} similarly. These are all proportional to the operator $Q_i[\lambda^l,[Q,\lambda^l]]_j$. Using the identity in Eq.\ (\ref{I_lambda'_lambda}), we find that
\begin{eqnarray}
\sum_lQ_{i'i}[\lambda^l,[Q,\lambda^l]]_{j'j}I_l &=& -4(2I_\pi+I_K)Q_{i'i}Q_{j'j} \nonumber \\
&& +4(I_\pi-I_K)Q_{i'i}\left[3(M^sQ)_{j'j} \right. \nonumber \\
\label{reducq[lql]}
&& \left. -\openone_{j'j}\textrm{Tr}\,M^sQ\right].
\end{eqnarray}
Adding the corresponding contributions with $j$ and $i$ interchanged and those from the other quark pairs and identifying the structures that appear, we obtain the effective operator
\begin{eqnarray}
&&  -8(2\mathcal{I}_{2,\pi}+\mathcal{I}_{2,K})\Gamma_4+24(\mathcal{I}_{2,\pi}-\mathcal{I}_{2,K})\Gamma_{13} \nonumber \\
\label{reducq[lql]_2}
&& \ \ -8(\mathcal{I}_{2,\pi}  -\mathcal{I}_{2,K})\mathrm{Tr}\,M^sQ\sum_iQ_i,
\end{eqnarray}
where $\mathrm{Tr}\,M^sQ=-1/3$ and
\begin{equation}
\label{I_(7)}
\mathcal{I}_{2,l} = I_{5,l}+I_{6,l}-I_{7,l}-I_{8,l}.
\end{equation}

The operator $\sum_iQ_i$ in the second line is not one of our standard set. However, we can rewrite it as
\begin{equation}
\label{reduc_sumQ}
\sum_iQ_i=Q_u\openone+\sum_i(Q_d-Q_u)M^d_i+\sum_i(Q_s-Q_d)M^s_i. 
\end{equation}
The first term in this expression leads to a change in the overall baryon mass parameter $m_B$ while the third term changes the unknown strange-quark parameter $\tilde{\alpha}_m$ in Eq.\ (\ref{deltaL_M}). Dropping these terms, we find that the diagrams in Figs.\ \ref{Fig7} give a contribution
\begin{eqnarray}
&& -8(2\mathcal{I}_{2,\pi}+\mathcal{I}_{2,K})\Gamma_4+24(\mathcal{I}_{2,\pi}-\mathcal{I}_{2,K})\Gamma_{13} \nonumber \\
\label{reducq[lql]_3}
&& \ -\frac{8}{3}(\mathcal{I}_{2,\pi}-\mathcal{I}_{2,K})(M^d_i+M^d_j+M^d_k)
\end{eqnarray}
to the effective mass Hamiltonian.

The total distinguishable contribution to the baryon masses from charge interactions follows from Eqs.\ (\ref{reduction1}), (\ref{reduc[q,lambda]^2_3}), and (\ref{reducq[lql]_3}):
\begin{eqnarray}
\mathcal{H}_{\textrm{charge}} &=& [I_{QQ}+6\mathcal{I}_{1,\pi}-8(2\mathcal{I}_{2,\pi}+\mathcal{I}_{2,K})]\Gamma_4  \nonumber \\
&& - 2\mathcal{I}_{1,\pi}(\Gamma_5-\Gamma_2) + \left[-12(\mathcal{I}_{1,\pi}-\mathcal{I}_{1,K}) 
 +24(\mathcal{I}_{2,\pi}-\mathcal{I}_{2,K})\right]\Gamma_{13} \nonumber \\
&& -2 (\mathcal{I}_{1,\pi}-\mathcal{I}_{1,K})(6\Gamma_{13} -3\Gamma_{10}-2\Gamma_{14}+\Gamma_{11}) \nonumber \\
\label{Hem}
&& +\left[2\mathcal{I}_{1,\pi}-\frac{8}{3}( \mathcal{I}_{2,\pi}-\mathcal{I}_{2,K})\right](M^d_i+M^d_j+M^d_k). \end{eqnarray}
%


\subsubsection{\label{subsubsec:magnetic} Corrections to the magnetic interactions}

The mesonic corrections to the moment-moment term depicted in Fig.\ \ref{Fig8} can be treated similarly. The corrections from the diagram in Fig.\ \ref{Fig8}\,(b), Eq (\ref{moment_exchange}), involve the structures $[\mu,\lambda^l]_i[\mu,\lambda^l]_j$ and $\{\mu,\lambda^l\}_i\{\mu,\lambda^l\}_j$. These can be reduced using the identity in Eq.\ (\ref{I_lambda'_lambda}) with the results
\begin{eqnarray}
\sum_l [\mu,\lambda^l]_{i'i}[\mu,\lambda^l]_{j'j}I_l
&=& 2I_\pi\left(2\mu_{i'j}\mu_{j'i}-\openone_{i'j}\mu^2_{j'i}-\mu^2_{i'j}\openone_{j'i}\right)
\nonumber \\
&& -2(I_\pi-I_K)\left[2\mu_{i'j}(\mu M^s)_{j'i}+2(\mu M^s)_{i'j}\mu_{j'i} \right. \nonumber \\
\label{reduc[mu,lambda]^2}
&& \left. - \mu^2_{i'j}M^s_{j'i} - M^s_{i'j}\mu^2_{j'i}-(\mu M^s \mu)_{i'j}
\openone_{j'i}-\openone_{i'j}(\mu M^s \mu)_{j'i}\right],
\end{eqnarray}
and
\begin{eqnarray}
\sum_l \{\mu,\lambda^l\}_i \{\mu,\lambda^l\}_j I_l
&=& 2I_\pi [ \mu^2_{i'j}\openone_{j'i} + \openone_{i'j}\mu^2_{j'i} +2 \mu_{i'j}\mu_{j'i}] -
4\left(I_\pi- I_\eta / 3 \right)\mu_{i'i}\mu_{j'j}  \nonumber \\
&-& 2\left(I_\pi-I_K\right) \left[\mu^2_{i'j}M^s_{j'i} + M^s_{i'j}\mu^2_{j'i}+
(\mu M^s \mu)_{i'j}\openone_{j'i}+ \openone_{i'j}(\mu M^s \mu)_{j'i} \right. \nonumber \\
&& + \left. 2\mu_{i'j}(\mu M^s)_{j'i} + 2(\mu M^s)_{i'j}\mu_{j'i}\right] \nonumber \\
&& +4(I_\pi-I_\eta)[\mu_{i'i}(M^s\mu)_{j'j}+(M^s\mu)_{i'i}\mu_{j'j}] \nonumber \\
\label{4t}
&& + 4\left(I_\pi-4I_K+3I_\eta\right)(\mu M^s)_{i'i}(M^s\mu)_{j'j}.  
\end{eqnarray}

These expressions appear multiplied by the common factor $\bsig_i\cdot\bsig_j$. A further rather lengthy reduction using the expression $\mu_i=\mu_aQ_i+\mu_b(QM^s)_i$ to display the charge dependence of the magnetic moments, and the exchange operator in Eq.\ (\ref{exchange_op}) to rearrange the skew indices, gives the contribution of Fig. 8\,(b) to the baryon mass differences as 
\begin{eqnarray}
\mathcal{H}_{9} & = & -6\mu_a^2I_{9,\pi}\Gamma_1 -\frac{2}{3}\mu_a^2 I_{9,\pi} (\Gamma_2 + 3 \Gamma_4)+
   6 \mu_a^2 (I_{9,\pi} - \frac{4}{27}I_{9,\eta}) \Gamma_5 \nonumber \\
&& + 2 \left[ (3\mu_a^2 + 4 \mu_a \mu_b
   + 2 \mu_b^2)I_{9,\pi} - 3 (\mu_a+\mu_b)^2 I_{9,K}\right]\Gamma_7 \nonumber \\
&&   + \frac{2}{3}\left[ \mu_a^2 I_{9,\pi} - (\mu_a+\mu_b)^2I_{9,K}\right] \Gamma_8 \nonumber \\
&& + \frac{2}{3}\mu_a^2 (I_{9,\pi} - I_{9,K}) (9 \Gamma_{10} + \Gamma_{11})
  + 4 \mu_a \left[ \mu_a I_{9,\pi} - (\mu_a+\mu_b)I_{9,K}\right] \Gamma_{13} \nonumber \\
&& + \frac{4}{3} \mu_a \left[ - 9 \mu_a I_{9,\pi} + 5(\mu_a+\mu_b)I_{9,K}
     + \frac{4}{3} (3\mu_a+2\mu_b) I_{9,\eta} \right] \Gamma_{14} \nonumber \\
&& + \frac{2}{3}\mu_b(2\mu_a+\mu_b) (I_{9,\pi} - I_{9,K})(9 \Gamma_{19}
   + \Gamma_{20}) \nonumber \\
&& - \frac{2}{3}\mu_b \left[ (2\mu_a+\mu_b) I_{9,\pi}
   + 6 (\mu_a+\mu_b)I_{9,K} \right] \Gamma_{25} \nonumber \\
&& - \frac{2}{3} \left[ (16\mu_a^2 + 26 \mu_a \mu_b + \mu_b^2)I_{9,\pi}
   - 2(10 \mu_a + 11 \mu_b)(\mu_a +\mu_b)I_{9,K} \right. \nonumber \\
\label{fig8b}
&& \left.   + \frac{4}{3}(3\mu_a^2 + 12 \mu_a \mu_b + 10 \mu_b^2) I_{9,\eta} \right]\Gamma_{26}.  
\end{eqnarray}

We will not need this full expression in our later analysis of mass splittings within isospin multiplets. In particular, the terms proportional to $\Gamma_7$ and $\Gamma_8$ can be absorbed in the unknown mass parameters $\tilde{\alpha}_m$ and $\tilde{\alpha}_{ss}$ defined in Eq. (2.37)  and can therefore be dropped in fits to data. In addition, the term in $\Gamma_1$ can be converted to the mass-difference form $\sum_i M^d_i$ using the result in Eq.\ (\ref{reduction_Gamma1}) and absorbing the extra pieces in the input parameters $m_B$ and $\tilde{\alpha}_m$.  

The terms in $\Gamma_{19}$, $\Gamma_{20}$, $\Gamma_{25}$, and $\Gamma_{26}$ complete the set of two-body operators given in Eqs.\ (\ref{2,4,5})-(\ref{25,26}). However, these last operators involve two factors of $M^s$ so act only in the $\Xi$, $\Xi^*$, and $\Omega$ systems where the $s$ quarks are necessarily in triplet spin configurations with $\bsig_i\cdot\bsig_j=1$. As a result, $\Gamma_{19}\equiv\Gamma_{20}$ and $\Gamma_{25}\equiv\Gamma_{26}$ for the octet and decuplet baryons. These terms do not affect the mass splittings within the  $\Xi$ and $\Xi^*$ multiplets and can be ignored in studying those splittings. 

The mass operator $\mathcal{H}_{10}$ obtained from the diagram in Fig.\ \ref{Fig8}\,(c) involves the new structures
\begin{eqnarray}
\sum_l \mu_{i'i} (\mu \lambda^l \lambda^l + \lambda^l \lambda^l \mu )_{j'j} I_l + (i \leftrightarrow j)
& = & 4(3 I_\pi + 2I_K + \frac{1}{3} I_\eta)\mu_{i'i} \mu_{j'j} - 2 (3 I_\pi - 2I_K - I_\eta) \nonumber \\
&& \times [\, \mu_{i'i} (\mu M^s)_{j'j} + (\mu M^s)_{i'i} \mu_{j'j} \,] ,  \\
\sum_l \mu_{i'i}[\lambda^l,[\mu,\lambda^l]]_{j'j} I_l + (i \leftrightarrow j)
& = & -8(2 I_\pi + I_K )\mu_{i'i} \mu_{j'j} \nonumber \\
&& +12(I_\pi - I_K)[\, \mu_{i'i} (\mu M^s)_{j'j} + (\mu M^s)_{i'i} \mu_{j'j} \,] \nonumber \\
&& - 4(I_\pi - I_K) {\rm Tr} (M^s \mu)[\,\mu_{i'i} \openone_{j'j}+\openone_{i'i} \mu_{j'j} \,] .
\end{eqnarray}
Substituting the expression in Eq.\ (\ref{QM_moments}) for the moments and identifying the resulting structures in Morpurgo's list, we find that the contribution of Fig. 8 \,(c) to the baryon mass differences is 
\begin{eqnarray}
\mathcal{H}_{10} & = & \frac{4}{3} \mu_a^2 \,(4 I_{9,\pi} +
3 I_{9,K} + \frac{2}{3}I_{9,\eta}) \,\Gamma_5 \nonumber \\
&& + \frac{4}{3} \mu_a \left[ (5 \mu_b-3\mu_a) I_{9,\pi} + (\mu_a +7 \mu_b)I_{9,K}
+ 2(\mu_a + \frac{5}{3}\mu_b)I_{9,\eta} \right] \,\Gamma_{14} \nonumber \\
&& + \frac{4}{3} \mu_b \left[ (\mu_b-3\mu_a) I_{9,\pi} + (\mu_a +4 \mu_b)I_{9,K}
+ 2(\mu_a + \frac{4}{3}\mu_b)I_{9,\eta} \right] \,\Gamma_{26} \nonumber \\
\label{fig8c}
&& -\frac{4}{3}(\mu_a + \mu_b) (I_{9,\pi} -I_{9,K})
\mathrm{Tr}\,M^sQ \left[ \mu_a \sum_iQ_i +  \mu_b \sum_i(QM^s)_i \right] .
\end{eqnarray}
where we have noted that $I_{10,l}=I_{9,l}$. 

The operator in the last line can be reduced to the form $\sum_iM^d_i$ by using the result in Eq.\ (\ref{reduc_sumQ}) and the identity $\sum_i(QM^s)_i=-(1/3)\sum_iM^s_i$ and then absorbing the pieces proportional to $Q_u\sum_i\openone_i$ and $\sum_iM^s_i$ in the parameters $m_B$ and $\tilde{\alpha}_m$,
\begin{eqnarray}
 && -\frac{4}{3}(\mu_a + \mu_b) (I_{9,\pi} -I_{9,K})
\mathrm{Tr}\,M^sQ \left[ \mu_a \sum_iQ_i +  \mu_b \sum_i(QM^s)_i \right] \nonumber \\
\label{reduce_trace}
&& \quad \rightarrow  \frac{4}{9}\mu_a(\mu_a + \mu_b) (I_{9,\pi} -I_{9,K})\sum_iM^d_i.
\end{eqnarray}

The total distinguishable contribution to the baryon mass differences from magnetic moment interactions is 
\begin{eqnarray}
\mathcal{H}_{\textrm{moment}} &=& I_{\mu\mu}(\mu_a^2\Gamma_5 + 2\mu_a\mu_b\Gamma_{14} + \mu_b^2\Gamma_{26}) + \mathcal{H}'_{9}+ \mathcal{H}'_{10} \nonumber \\
\label{H_moment}
&& + \left[2\mu_a^2I_{9,\pi}+\frac{4}{9}\mu_a(\mu_a+\mu_b)\left(I_{9,\pi}-I_{9,K}\right)\right]\left(M^d_i + M^d_j +M^d_k\right),
\end{eqnarray}
where $\mathcal{H}'_{9}$ is the operator obtained by dropping the terms  in $\mathcal{H}_{9}$, Eq.\ (\ref{fig8b}) proportional to $\Gamma_1$, $\Gamma_7$, and $\Gamma_8$, and  $\mathcal{H}_{10}$ is the operator obtained by dropping the trace term in the last row in Eq.\ (\ref{fig8c}).


\subsubsection{\label{mass_diff} Corrections from mass differences}

The effects of the $d,\,u$ quark mass differences on the baryon masses and on the single meson exchange amplitude contribute a further non-electromagnetic term that affects the mass splittings within isospin multiplets. From Eqs.\ (\ref{deltam_du}) and (\ref{meson_mass2}), with the indices in the latter rearranged,
\begin{eqnarray}
\mathcal{H}_{du} &=& \Delta_{du}(M^d_i+M^d_j+M^d_k)  \nonumber \\
\label{Hdu}
&& -\frac{2}{3} \Delta^ M_q I_{4,K^0}\,\left[M^u_iM^s_j+M^s_jM^u_i + (j,k)+(k,i)\right](3-  \bsig_i\cdot\bsig_j).
\end{eqnarray}

The term $\Delta_{du}\sum_iM^d_i$ in this expression arises from the difference between the $d$- and $u$-quark masses.  There are additional contributions to the baryon masses from $\mathcal{H}_{\textrm{charge}}$ and $\mathcal{H}_{\textrm{moment}}$ that have the same structure but arise from mesonic corrections to electromagnetic interactions. If we assume that these are calculable, their contributions can be subtracted from the overall coefficient of $\sum_iM^d_i$ obtained in fits to the data to determine $\Delta_{du}$. Since $\Delta_{du}$ can also be estimated as in Eq.\ (\ref{d-u/s-u}), this procedure provides a test of the theory sensitive to both quark-mass and electromagnetic effects.

We note finally that the second term in Eq.\ (\ref{Hdu}) can be expressed in terms of the $\Gamma$s using the identity
\begin{eqnarray}
[M^u_iM^s_j+M^s_iM^u_j +\cdots](3-\bsig_i\cdot\bsig_j ) &=& -\left[2Q_i(QM^s)_j+2(QM^s)_iQ_j  -Q^2_iM^s_j-M^s_iQ^2_j\right.\nonumber \\
\label{M^uM^s}
&& \left.-(QM^sQ)_i\openone_j-\openone_i(QM^sQ)_j\right](3-\bsig_i\cdot\bsig_j)+ \cdots \nonumber  \\
&=& -12\Gamma_{13}+6\Gamma_{10}+4\Gamma_{14}-2\Gamma_{11}. 
\end{eqnarray}
%


\subsection{\label{subsec:fit} Fit to the data on mass splittings}

The contributions of the various one- and two-body charge and spin structures $\Gamma_i$ to the baryon mass splittings can be determined from results given in \cite{Morpurgo_EM1}, especially Table III of that reference. The contributions of the relevant operators are summarized in Table\ \ref{table:weights_for_Gammas}.
\begin{table*}[h]
\caption{\label{table:weights_for_Gammas}Table of the contributions of the operators $\Gamma_i$ to the mass splittings within baryon multiplets.  The one- and two-body operators that are not listed either give contributions that can be absorbed in the input mass parameters ($\Gamma_1$, $\Gamma_7$,  $\Gamma_8$) or are identical  for the baryon octet and decuplet to others on the list ($\Gamma_{20}\equiv\Gamma_{19}$, $\Gamma_{26}\equiv\Gamma_{25}$.}
\begin{ruledtabular}
\begin{tabular}{lcrrrrrrrrr}
& $\sum_iM^d_i$ & $\Gamma_2$ & $\Gamma_4$ & $\Gamma_5$ & $\Gamma_{10}$ & $\Gamma_{11}$ & $\Gamma_{13}$ & $\Gamma_{14}$ & $\Gamma_{19}$ & $\Gamma_{25}$ \\[0.5ex]
\hline
$n-p$ & 1 & $-1/3$ & -1/3 & -1/3 & 0 & 0 & 0 & 0 & 0 & 0 \\
$\Sigma^--\Sigma^0$ & 1  & 1/6 & 2/3 & -1/3 & -1/3 & -1/3 & 1/3 & 1/3 & 0 &0 \\
$\Sigma^0-\Sigma^+$ & 1 & 1/6 & -1/3 &-4/3 & -1/3 & -1/3 & 1/3 & 1/3 & 0 & 0 \\
$\Xi^--\Xi^0$ & 1 & 2/3 & 2/3 & -4/3 & -2/3 & -2/3 & 2/3 & 2/3 & 0 &0 \\
$\Delta^--\Delta^0$ & 1 & -1/3 & 2/3 & 2/3 & 0 & 0 & 0 & 0 & 0 & 0 \\
$\Delta^0-\Delta^+$ & 1 &-1/3 & -1/3 & -1/3 & 0 & 0 & 0 & 0 & 0 & 0 \\
$\Delta^+-\Delta^{++} $ & 1 & -1/3 & -4/3 & -4/3 & 0 & 0 & 0 & 0 & 0 & 0 \\
$\Sigma^{*-}-\Sigma^{*0}$ & 1 & -1/3 & 2/3 & 2/3 & -1/3 & -1/3 & 1/3 & 1/3 & 0 & 0 \\
$\Sigma^{*0}-\Sigma^{*+}$ & 1 & -1/3 & -1/3 & -1/3 & -1/3 & -1/3 & 1/3 & 1/3 & 0 & 0 \\
$\Xi^{*-}-\Xi^{*0}$ & 1 &  -1/3 & 2/3 & 2/3 & -2/3 & -2/3 &2/3 & 2/3 & 0 & 0
\end{tabular}
\end{ruledtabular}
\end{table*}

As may be seen from Table\ \ref{table:weights_for_Gammas}, $\Gamma_{10}=\Gamma_{11}=-\Gamma_{13}=-\Gamma_{14}$ when restricted to mass differences, though the operators are themselves distinct.  In addition, $\Gamma_{19}=\Gamma_{20}$ and $\Gamma_{25}=\Gamma_{26}$ do not contribute. The last five columns in the table can therefore be eliminated. To determine if the remaining operators are independent, we consider the $5\times 5$ Gram matrix $\mathcal{M}_{\Gamma}=\mathcal{M}^T\mathcal{M}$  associated with  the $10\times 5$ matrix $\mathcal{M}$ defined by the first five columns in the table. This is a matrix of inner products of the remaining $\Gamma$s regarded as column vectors.  $\mathcal{M}_{\Gamma}$ has a vanishing determinant and one zero eigenvalue indicating that there is one relation among the  five operators.  We easily find  that $\Gamma_2=-\frac{1}{3}\sum_iM^d_i+\frac{1}{2}(\Gamma_4-\Gamma_5)$ when restricted to the space of mass differences, giving an extra relation not immediately evident in the table. As a result, we can bring the mass difference operator to the form
\begin{equation}
\label{H1}
\mathcal{H}_{\textrm{em}} = a\sum_iM^d_i+b\Gamma_4+c\Gamma_5+d\Gamma_{13}.
\end{equation}
The choice of the independent operators is natural. The first three are introduced by the quark mass corrections and the Coulomb and magnetic moment interactions between quarks independently of any mesonic corrections so are natural choices for independent operators, while $\Gamma_{13}$ enters prominently in the mesonic corrections. The coefficients in this expression follow from the results above.

Since there are only four independent parameters in $\mathcal{H}_{\textrm{em}}$, we expect there to be
six linear relations among the ten mass differences.\footnote{To see this formally, we note that  the $10\times 10$ Gram matrix $\mathcal{M}_G=\mathcal{M}\mathcal{M}^T$ for the mass differences  has six  zero eigenvalues. The corresponding eigenvectors give six null relations or sum rules for the mass differences.} These are the well-known sum rules 
\begin{eqnarray}
&& \Delta^0-\Delta^+ = n-p \nonumber \\
&& \Delta^--\Delta^{++} = 3(n-p) \nonumber \\
&& \Delta^0-\Delta^{++} = 2(n-p) + (\Sigma^0-\Sigma^+) - (\Sigma^--\Sigma^0) \nonumber \\
\label{sum_rules}
&& \Xi^--\Xi^0= (\Sigma^--\Sigma^+)-(n-p) \\
&& \Xi^{*-}-\Xi^{*0} = (\Sigma^{*-}-\Sigma^{*+})-(n-p) \nonumber \\
&& 2 \Sigma^{*0}-\Sigma^{*+} -\Sigma^{*-}= 2\Sigma^0-\Sigma^+-\Sigma^-. \nonumber
\end{eqnarray}
The fourth is the Coleman-Glashow relation, suggested originally on the basis of  an unbroken SU(3) flavor symmetry \cite{Coleman-Glashow}. All the sum rules were later established for nonrelativistic quark models with only one- and two-body interactions independently of the flavor symmetry breaking \cite{Rubenstein1,Rubenstein2,Ishida,Gal-Scheck}.  The sum rules are first violated by three-body terms.  

Since the nonrelativistic quark model has the same spin-flavor structure as the relativistic effective field theory and can be regarded as a parametrization of the latter \cite{DHJ1,Morpurgo_QM1,Morpurgo_QM2,Morpurgo_QM3}, the sum rules continue to hold in the relativistic case as a consequence of the structure imposed by QCD. We have seen that there are no three-body terms even when we include the leading mesonic corrections in the chiral effective field theory, that is, when the baryon mass corrections are calculated to two loops overall. The first corrections enter at three loops. We therefore expect the sum rules to hold with reasonable accuracy. 

The $\Delta$ baryon masses are not determined with sufficient accuracy for the first three sum rules to give a real test of this expectation. The results for the remaining three as written are, in order, $6.48\pm 0.24\ \textrm{MeV}=6.79\pm 0.08\ \textrm{MeV}$, $3.20\pm 0.68\ \textrm{MeV}=3.11\pm 0.64\ \textrm{MeV}$, and $-2.60\pm 1.18\ \textrm{MeV}=-1.535\pm 0.08\ \textrm{MeV}$. If we transfer all the terms to the left hand sides of the equations, the results for these sum rules are $-0.31\pm 0.25\ \textrm{MeV}$, $0.09\pm 0.93\ \textrm{MeV}$, and $-1.06\pm 1.18\ \textrm{MeV}$, all equal to zero within the experimental uncertainties. No significant violations of the sum rules are evident.

A fit to the mass splittings other than those for the $\Delta$ baryons is given in Table \ref{table:fit_to_data}. The overall fit is good with a average deviation from experiment of 0.12 MeV and a $\chi^2$ per degree of freedom of 0.99. Note that the sum rules in Eq.\ (\ref{sum_rules}) restrict what can be done in fitting the data. We cannot, for example, fit the four mass differences within the baryon octet exactly using the four parameters in Eq.\ (\ref{H1}) because of the Coleman-Glashow relation: $\Gamma_{13}=(\Gamma_4-\Gamma_5)/3$ when restricted to this sector so only three of the parameters are independent. 
\begin{table}[h]
\caption{\label{table:fit_to_data}A weighted fit to the seven accurately known baryon
mass splittings using the expressions in Eq.\ (\protect{\ref{H1}}) with the coefficients
given in Table \protect{\ref{table:weights_for_Gammas}}. A best fit is
obtained at the values (in MeV) of $a = 1.88 \pm 0.01$, $b = 3.52 \pm 0.02$,
$c =-1.77 \pm 0.00$, and $d = 0.22 \pm 0.03$. The average deviation of the
fit from experiment is $0.12$ MeV. The experimental data are from \cite{PDG}.}
\begin{ruledtabular}
\begin{tabular}{lrl}
Splittings & Calculated \, & Experiment \\
\hline
$n - p$ & 1.29 $\pm$ 0.01 & 1.293 $\pm$ 0.000 \\
$\Sigma^- - \Sigma^+$ & 8.03 $\pm$ 0.03 &  8.08 $\pm$ 0.08  \\
$\Sigma^- - \Sigma^0$ & 4.89 $\pm$ 0.02  &  4.807 $\pm$ 0.035\\
$\Xi^- - \Xi^0$ & 6.74 $\pm$ 0.02& 6.48 $\pm$ 0.24 \\
$\Sigma^{*-} - \Sigma^{*+}$ & 4.49 $\pm$ 0.03 & 4.40 $\pm$ 0.64 \\
$\Sigma^{*-} - \Sigma^{*0}$ & 3.12 $\pm$ 0.02 & 3.50 $\pm$ 1.12 \\
$\Xi^{*-} - \Xi^{*0}$ & 3.19 $\pm$ 0.02 & 3.20 $\pm$ 0.68  \\
$\Delta^{++} - \Delta^0$ & -0.84 $\pm$ 0.03 &  --- \\
$\Delta^{++} - \Delta^-$ & -3.88 $\pm$ 0.03 & --- \\
$\Delta^+ - \Delta^0$ & -4.34 $\pm$ 0.02 & --- \\
\end{tabular}
\end{ruledtabular}
\end{table}

The remaining question for the present analysis is whether the effects of the quark masses, the Coulomb and moment-moment interactions, and the mesonic corrections can account for the parameters in the fit. We will not examine this in detail here but will estimate the principal  contributions in the following section. One of us (PH) is calculating the corrections and will report separately on the results of his analysis.



\subsection{\label{subsec:matrixelements} Estimates of matrix elements}

To get an idea of the likely size of various electromagnetic contributions to the baryon mass splittings, we have made some estimates of the relevant matrix elements which we report here. We start with the basic Coulomb interaction term $\mathcal{H}_{QQ}$ which we write in the form in Eq.\ (\ref{Coulomb_energy2}). To evaluate the matrix elements of this operator, we need information on the internal structure of the baryons. This is equivalent  in the effective field theory approach to adding further momentum structure, but relatively soft structure corresponding to extended spatial wave functions as in semirelativistic dynamical models. The structure quarks $q_i$, which have so far described only the spin and flavor structure of the baryon, then become dynamical, but still act effectively as dressed rather than QCD quarks. 

The semirelativistic theory of baryon structure has been considered by a number of authors and is quite successful. See, for example, Brambilla \textit{et al.} \cite{Brambilla} and Carlson \textit{et al.} \cite{Carlson} for the theoretical background and Capstick and Isgur \cite{Capstick} and the extensive references therein for applications. For simplicity, we will use the model considered in \cite{DH_moments1} in which the baryon masses are calculated variationally for the semirelativistic Hamiltonian of Brambilla \textit{et al.} using Gaussian wave functions. The results agree with those of a similar calculation by Carlson, Kogut, and Pandharipande \cite{Carlson} and are consistent with those of the much more extensive calculations of Capstick and Isgur \cite{Capstick}. 

We will use Jacobi coordinates to describe the positions of the quarks. Define
\begin{eqnarray}
{\rb}_{ij} &=& \bx_i - \bx_j \, , \ \
{\bf R}_{ij} = \frac{m_i\bx_i + m_j\bx_j}{m_{ij}} \, , \nonumber \\
\label{Jacobi_space}
{\rb}_{ij,k} &=& {\bf R}_{ij} - \bx_k \, = \,
\frac {m_i(\bx_i - \bx_k) + m_j(\bx_j - \bx_k)}{m_{ij}} \, , \nonumber \\
{\bf R}_{ijk} &=& \frac {m_{ij}{\bf R}_{ij} + m_k\bx_k}{M} ,
\end{eqnarray}
where the $\bx_i$ are the particle coordinates, $m_{ij}=m_i+m_j$, $M=m_i+m_j+m_k$, and ${\bf R}_{ijk}$ is the usual center-of-mass coordinate. The roles of $i$, $j$, and $k$ are completely symmetric at this stage. However, it is reasonable to neglect the very small difference between the effective masses of the $u$ and $d$ quarks in the dynamical calculations. At least two of the quarks in each baryon are then identical or have the same mass. We label these 1 and 2, with the odd quark labelled 3. We then define the internal Jacobi coordinates  $\brho$ and $\blam$ as $\brho = {\rb}_{12}$ and $\blam = {\rb}_{12,3}$. Alternatively, we can use coordinates with the role of $(1,2)$ replaced by $(2,3)$ or $(3,1)$ in the definition, and define
$\brho' = {\rb}_{23}$,
$\blam' = {\rb}_{23,1}$, or
$\brho'' = {\rb}_{31}$, 
$\blam'' = {\rb}_{31,2}$.
The coordinate pairs $\brho'$, $\blam'$ and $\brho''$, $\blam''$ can be expressed in terms of
$\brho$ and $\blam$ and conversely, so one can work with whichever of the pairs is most convenient and switch between them as necessary. The spatial volume element is simply $d^3R\,d^3\rho\,d^3\lambda$, and equivalently for the other pairs of internal coordinates.

We may also use the momentum coordinates
\begin{eqnarray}
{\pb}_{ij} &=& \frac{m_j{\pb}_i - m_i{\pb}_j}{m_{ij}}  \, , \ \ \
{\bf P}_{ij} = {\pb}_i + {\pb}_j  \, , \nonumber \\
{\pb}_{ij,k} &=& \frac{m_k{\bf P}_{ij} - m_{ij}{\pb}_k}{M} \, , \ \
{\bf P}_{ijk} = {\bf P}_{ij} + {\pb}_k  \, ,
\end{eqnarray}
where ${\bf P}_{ijk}\equiv{\bf P}$ is the total momentum. The Jacobi momentum coordinates are then the pairs
${\pb}_\rho = {\pb}_{12}$ and ${\pb}_\lambda = {\pb}_{12,3}$, or
${\pb}_{\rho '} = {\pb}_{23}$ and ${\pb}_{\lambda '} = {\pb}_{23,1}$, or
${\pb}_{\rho ''} = {\pb}_{31}$ and ${\pb}_{\lambda ''} = {\pb}_{31,2}$. One can choose to work with any of the pairs as there are linear relations among them. The volume element
in momentum space is $d^3P\,d^3p_\rho\,d^3p_\lambda$, and equivalently for the
other pairs of internal momenta.

With these definitions, the simplest versions of the position-space variational wave functions in \cite{DH_moments1} for the $L=0$ ground states are just the Gaussians\footnote{We actually considered
somewhat more general functions, including terms with angular excitations and the with the Gaussians multiplied by polynomials, but the basic results changed rather little. The simple Gaussians are sufficient to
illustrate the main features of the corrections.}
\begin{equation}
\label{gausswf}
\psi_0(\brho,\blam) = \left( \frac {\beta_\rho \beta_\lambda }{ \pi} \right)^{3/2}
 \exp{ [ -\frac{1}{2}(\beta_\rho^2 \rho^2 + \beta_\lambda^2 \lambda^2)]},
\end{equation}
equivalent to the momentum-space functions
\begin{equation}
\label{momentum_wf}
\tilde{\psi}_0({\pb}_\rho,{\pb}_\lambda) =
\left(\frac{1}{\pi\beta_\rho\beta_\lambda}\right)^{3/2}
\exp{\left(-\frac{p_\rho^2}{2\beta_\rho^2}-\frac{p_\lambda^2}{2\beta_\lambda^2} \right)}.
\end{equation}
The variational parameters $\beta_\rho$ and $\beta_\lambda$ differ slightly for the different baryons, reflecting the effects of the differing quark masses on the wave functions \cite{DH_moments1}.

It is straightforward to calculate the Coulomb integrals in $\mathcal{H}_{QQ}$ using the identifications $|\bx_1-\bx_2|=\rho$, $|\bx_2-\bx_3|=\rho'$, and $|\bx_3-\bx_1|=\rho''$ and changing coordinates appropriately \cite{DH_moments1}. The result is
\begin{equation}
\label{<Coulomb_energy3>}
\mathcal{H}_{QQ}^B = \frac{2\alpha_{\textrm{em}}}{\sqrt{\pi}}\left[Q_1Q_2\beta_\rho^B+(Q_2Q_3+Q_3Q_1)\beta_\lambda^{'\,B}\right],
\end{equation}
where $\beta_\lambda^{'2}=\beta_\lambda^2/(1+x/4)$, $x=\beta_\lambda^2/\beta_\rho^2$, and $B$ labels the baryon in question. 

This structure departs from the simple proportionality to $\Gamma_4=\sum[Q_iQ_j]$ in Eqs.\ (\ref{Coulomb_energy}) and (\ref{reduction1}) because of the dependence of the parameters on the baryon. The changes introduce small terms in $\Gamma_{13}$ and $\Gamma_{25}$, and more interestingly, a small three-body term proportional to $\Gamma_{16}$ reflecting the influence of a massive strange quark on the correlations between the remaining quarks. However, this change  contributes only $\sim 0.05$ MeV to the discrepancy in the Coleman-Glashow sum rule. It is sufficient for our purposes to ignore the baryon dependence of $\beta_\rho$ and $\beta_\lambda$ and simply use the values $\beta_\rho=0.340$ GeV and $\beta_\lambda=\sqrt{4/3}\,\beta_\rho=0.393$ GeV obtained for the nucleon. The result is
\begin{equation}
\label{Coulomb_energy4}
\mathcal{H}_{QQ} = \frac{2\alpha_{\textrm{em}}}{\sqrt{\pi}}\beta_\rho\Gamma_4 = (2.80\ \textrm{MeV})\,\Gamma_4.
\end{equation}
Given the extreme simplicity of the wave function, this contribution is in reasonable agreement with the term $b\Gamma_4$ in $\mathcal{H}_{\textrm{em}}$, Eq.\ (\ref{H1}), obtained in fitting the data on mass splittings, $b=3.52$ MeV. The more flexible wave function used by Carlson \textit{et al.} gives a larger Coulomb energy corresponding to $b\approx 3.3$ MeV.

The leading contribution to the $\Gamma_5$ term in $\mathcal{H}_{\textrm{em}}$ is presumably the part of the magnetic moment interaction proportional to $\mu_a^2$, Eq.\ (\ref{reduction1}). Evaluating $I_{\mu\mu}$ using Eq.\ (\ref{H_mumu2}) and the Gaussian wave functions, we get
\begin{equation}
\label{magnetic_energy}
\mathcal{H}_{\mu\mu} = -\frac{8\pi}{3}\frac{\beta_\rho^3}{\pi^{3/2}}\mu_a^2\,\Gamma_5+\cdots = -(0.953\ \textrm{MeV})\,\Gamma_5+\cdots.
\end{equation}
The coefficient of $\Gamma_5$ is somewhat small compared to that found in our fit, $c=-1.77$ MeV, but the sign is correct. As emphasized by Capstick and Isgur \cite{Capstick}, the magnitude is sensitive to short-distance correlations and is generally underestimated in the perturbative calculation.

The integrals associated with the mesonic corrections bring in other features. We will consider $I_{1,l}$, Eq.\ (\ref{I1l}), as an example. This integral factors into the product of a Coulomb integral and a mesonic integral. 
\begin{equation}
\label{I1l_2}
I_{1,l} = e^2\int\frac{d^3k}{(2\pi)^3|\bk|^2}\times\frac{1}{4f^2}\int\frac{d^3k'}{(2\pi)^32E_l'}\frac{k^{'2}}{E_l^{'2}}.
\end{equation}

This structure is easy to interpret.  As indicated by the time-ordered diagrams in Fig.\ \ref{Fig4}\,(a), the physical process consists of the emission of a meson followed by a Coulomb interaction in the intermediate baryon state and the subsequent reabsorption of the meson to reach the final baryon state. In the heavy-baryon approximation, the only allowed intermediate baryons are members of the ground state octet and decuplet. Excited states are substantially higher in mass and their contributions can be neglected. 

The dominant contributions to the spatial wave functions for the ground-state octet and decuplet have $L=0$ and are the same up to very small corrections that arise through the different contributions of higher orbital angular momenta \cite{DH_moments1}. If we neglect these corrections, the wave functions are the same for a given quark content, the usual quark-model picture, and the Coulomb matrix element is the same for either multiplet. If we neglect in addition the changes in the wave functions associated with changes in the quark masses, the emission and absorption matrix elements are the also same for octet-octet, octet-decuplet, and decuplet-decuplet transitions. 

Finally, when we include the internal structure of the baryon, the original plane-wave matrix element  $-(i/2f)\bsig_i\cdot\bk'$ for the emission of a meson by quark $i$ becomes
\begin{eqnarray}
&& \int d^3\rho\, d^3\lambda\, \psi^*(\brho,\blam) \left(-\frac{1}{2f}\bsig_i\cdot\nabla_ie^{i\bk'\cdot\bm{r}_i}\right)\psi(\brho,\blam) \nonumber \\
\label{axial1}
&& \quad =- \frac{i}{2f}\bsig_i\cdot\bk'\int d^3\rho\, d^3\lambda\, \left|\psi(\brho,\blam)\right|^2e^{i\bk'\cdot(\blam+\brho/2)} \nonumber \\
&& \quad =- \frac{i}{2f}\bsig_i\cdot\bk'F_A(k^{'2}),
\end{eqnarray}
where $F_A(k^{'2})$ is just the axial vector form factor of the baryon, and the mesonic factor in $I_{1.l}$ becomes
\begin{equation}
\label{I1l_3}
I_{1,l}' =\frac{1}{4f^2}\int\frac{d^3k'}{(2\pi)^32E_l'}\frac{k^{'2}}{E_l^{'2}}F_A^2(k^{'2}).
\end{equation}

In the case of the Gaussian wave functions discussed above, neglecting mass effects, $F_A(k^{'2})= \exp(-k^{'2}/4\beta_\rho^2)$. The form factor $F(k^{'2})=\Lambda^4/(\Lambda^2+k^{'2})^2$ with $\Lambda=930$ MeV used in our earlier analyses of baryon masses \cite{DH_masses,DHJ1} and magnetic moments \cite{DH_moments2,DH_moments3} is probably more realistic. This was modeled after the nucleon electromagnetic form factors and includes somewhat higher momenta. However, the two agree well for small values of $k^{'2}$.

Because the pion mass is small compared to the cutoff momenta, we can estimate the integral in Eq.\ (\ref{I1l_3}) by setting $M_\pi=0$. The integrals can then be done analytically for either form factor. The Gaussian form factor gives  $I_{1,l}'=\beta_\rho^2/16\pi^2f^2=0.085$, while the second form gives $I_{1,l}'=\Lambda^2/96\pi^2f^2=0.106$. Thus, $I_{1,\pi}\approx( 0.1)I_{QQ}$. This leads to 20\% corrections to the coefficient  of $\Gamma_4$ in $\mathcal{H}_{\textrm{charge}}$, Eq.\ (\ref{Hem}), and the coefficient $b$  in the full effective Hamiltonian  $\mathcal{H}_{\textrm{em}}$, Eq.\ (\ref{H1}). 

Similar methods can be used to estimate the corrections  associated with other diagrams. These appear to be of similar magnitude, and a full calculation is needed to establish how well the dynamical theory describes the coefficients in $\mathcal{H}_{\textrm{em}}$.


\section{\label{sec:conclusions} Conclusions}

Our results here consist of a thorough analysis of the electromagnetic contributions to the baryon masses including the first mesonic corrections to the basic electromagnetic terms. The analysis was done using the heavy-baryon effective field theory methods developed  in earlier work which connect naturally to the general parametrization of the electromagnetic effects given by Morpurgo \cite{Morpurgo_EM1}. 

We find that the electromagnetic corrections are purely two-body when calculated through one loop in the mesonic corrections, that is, to two loops overall. The contributions from diagrams that involve three quark flavor labels all vanish. As a result, the six sum rules among the ten octet and decuplet mass splittings derived many years ago in the nonrelativistic quark model \cite{Rubenstein1,Rubenstein2,Ishida,Gal-Scheck} continue to hold through two loops in the relativistic chiral effective field theory. 

The first corrections to the sum rules necessarily involve three-body effects, so at least two meson loops in addition to the electromagnetic interaction. This suggests strongly that the corrections to the well-satisfied Coleman-Glashow relation and the other sum rules in Eq.\ (\ref{sum_rules}) should be quite small, in agreement with arguments directly from QCD \cite{Morpurgo_hierarchy} and from the $1/N_c$ expansion \cite{Jenkins4,Jenkins5}. An estimate of a typical nonvanishing three body term in fact gives a value $\approx I^{'2}_{1,\pi}I_{QQ}\approx 0.01I_{QQ}$ where $I'_{1,l}$ is the integral in Eq.\ (\ref{I1l_3}) and $I_{QQ}$ is the Coulomb integral in Eq.\ (\ref{Coulomb_energy}), but the coefficients from the spin and flavor factors are not known and could be large enough to make the corrections significant given the typical coefficients in $\mathcal{H}_{\textrm{charge}}$, Eq.\ (\ref{Hem}).

It remains to determine the extent to which the mesonic and mass corrections to the basic electromagnetic interactions account numerically for the pattern of coefficients in $\mathcal{H}_{\textrm{em}}$. This will be investigated elsewhere. 

We note finally that the results of the present work can be combined with those in \cite{DHJ2,DH_masses} to obtain a complete description of the baryon masses, including the intermultiplet splittings, through one loop in the mesonic corrections in heavy-baryon chiral effective field theory. It is necessary in that application to start with the full expressions for the electromagnetic corrections since some $\Gamma$s which are equivalent for the splittings within multiplets are distinct in the general setting. Because there are still no three-body corrections \cite{DHJ2}, the nine two-body sum rules derived by Rubenstein \textit{et al.} \cite{Rubenstein2} continue to hold, and the octet and decuplet masses can be parametrized in terms of an overall mass $m_B$ and the eight distinct parameters in Eqs.\ (\ref{deltaL_M}) and (\ref{H1}). For recent discussions of these sum rules from the points of view of the quark model and the $1/N_c$ expansion, see Rosner \cite{Rosner} and Jenkins and Lebed \cite{Jenkins4,Jenkins5}.

\begin{acknowledgments}
One of the authors (LD) would like to thank the Aspen Center for Physics for its hospitality while parts of this work were done. The other author (PH) is grateful to the Department of Physics and Astronomy,
Indiana University South Bend, for its hospitality and support of work done there.
\end{acknowledgments}

\bibliography{EM_Bib.bib}

\end{document}